\documentclass[aip, reprint]{revtex4-2}

\usepackage{amsmath}
\usepackage{amssymb}
\usepackage{bm}  
\usepackage{bbm}  
\usepackage{booktabs}  
\usepackage{dcolumn}  
\usepackage{graphicx}  
\usepackage[mathlines]{lineno}  
\usepackage{makecell}  
\usepackage{multirow}  
\usepackage[normalem]{ulem}
\usepackage{physics}  
\usepackage{tabularx}  
\usepackage{upgreek}  
\usepackage{xcolor}  
\usepackage[pdfencoding=auto]{hyperref}  
\usepackage{cleveref}  
\usepackage{appendix}

\newcommand{\etal}{et.~al.}  
\newcommand{\boltzmann}{k_{\scriptscriptstyle\textrm{B}}}  
\newcommand{\inftes}[1]{\,\textrm{d}#1\,}  
\newcommand{\bjerrum}{\uplambda_\textrm{B}}  
\newcommand{\specA}{\nu}  
\newcommand{\specB}{\nu^{\prime}}  
\newcommand{\specC}{i}  
\newcommand{\specD}{\mu}  
\newcommand{\funcdepend}{[\{\rho_{\specC}\}]}  
\newcommand{\DADB}{d_{\specA}+d_{\specB}}  
\newcommand{\DADBTwo}{\tfrac{d_{\specA}+d_{\specB}}{2}}  
\newcommand{\nodc}[1]{\multicolumn{1}{c}{#1}}  
\newcommand{\nodcrl}[1]{\multicolumn{1}{c|}{#1}}  
\newcommand{\functionalmf}{\mathcal{F}_{\textrm{ex}}^{\textrm{MF}}}
\newcommand{\functionaldelta}{\mathcal{F}_{\textrm{ex}}^{\delta}}
\newcommand{\functionaltheta}{\mathcal{F}_{\textrm{ex}}^{\theta}}
\let\oldvec\vec
\renewcommand{\vec}[1]{\oldvec{\textrm{#1}}}  

\renewcommand{\epsilon}{\varepsilon}  
\let\oldint\int
\renewcommand{\int}{\oldint\limits}  
\let\oldiint\iint
\renewcommand{\iint}{\oldiint\limits}  

\newcommand{\orcid}[1]{\href{https://orcid.org/#1}{\includegraphics[height=\fontcharht\font`\B]{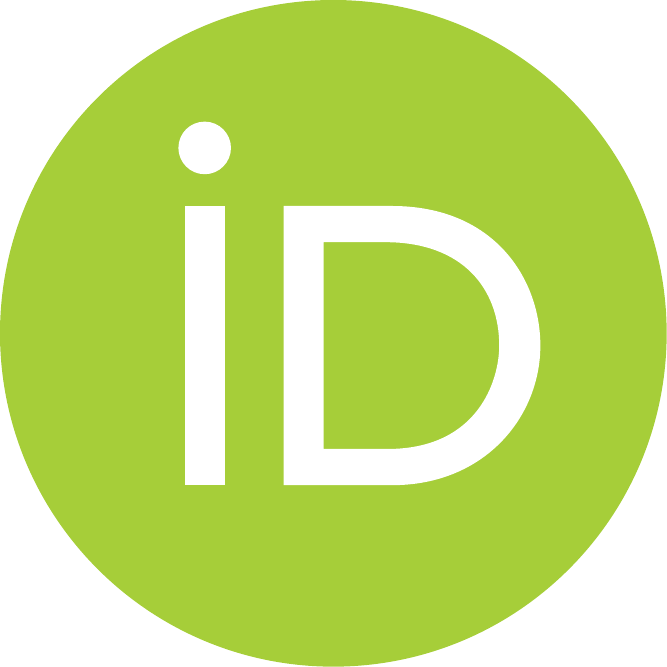}}}

\newcommand{\DFT}{DFT}
\newcommand{\lhs}{lhs}
\newcommand{\rhs}{rhs}
\newcommand{\MSA}{MSA}
\newcommand{\EDL}{EDL}
\newcommand{\FMT}{FMT}
\newcommand{\OZ}{OZ}
\newcommand{\PM}{PM}

\newcommand{\CDT}{CDT}
\newcommand{\MD}{MD}

\begin{document}

\title{The Primitive Model in Classical Density Functional Theory:\\Beyond the Standard Mean-Field Approximation}

\author{Moritz B\"ultmann~\orcid{0000-0002-4522-1849}\phantom{x}}
\email{moritz.bueltmann@physik.uni-freiburg.de}
\affiliation{Physikalisches Institut, Albert-Ludwigs-Universität,
    79104 Freiburg, Germany}
\author{Andreas H\"artel~\orcid{0000-0002-1352-2559}\phantom{x}}
\email{andreas.haertel@physik.uni-freiburg.de}
\affiliation{Physikalisches Institut, Albert-Ludwigs-Universität,
    79104 Freiburg, Germany}
\date{\today}

\begin{abstract}
\textbf{A peer-reviewed version of this article was published in J. Phys.: Condens. Matter 34 235101 on the 6th of April 2022~\href{https://doi.org/10.1088/1361-648x/ac5e7a}{https://doi.org/10.1088/1361-648x/ac5e7a}.}\\

The primitive model describes ions by point charges with an additional hard-core interaction.
In classical density-functional theory the mean-field electrostatic contribution can be obtained from the first order of a functional perturbation of the pair potential for an uncharged reference system of hard spheres.
This mean-field electrostatic term particularly contributes at particle separations that are forbidden due to hard-core overlap.
In this work we modify the mean-field contribution such that the pair potential is constant for distances smaller than the contact distance of the ions. We motivate our modification by the underlying splitting of the potential, which is similar to the splitting of the Weeks-Chandler-Andersen potential and leads to higher-order terms in the respective expansion of the functional around the reference system. 
The resulting formalism involves weighted densities similar to the ones found in fundamental measure theory.
To test our modifications, we analyze and compare density profiles, direct and total correlation functions, and the thermodynamic consistency of the functional via a widely established sum rule and the virial pressure formula for our modified functional, for established functionals, and for data from computer simulations.
We found that our modifications clearly show improvements compared to the standard mean-field functional, especially when predicting layering effects and direct correlation functions in high concentration scenarios; for the latter we also find improved consistency when calculated via different thermodynamic routes.
In conclusion, we demonstrate how modifications towards higher order corrections beyond mean-field functionals can be made and how they perform, by this providing a basis for systematic future improvements in classical density-functional theory for the description of electrostatic interactions.\end{abstract}

\maketitle

\section{Introduction}\label{sec:introduction}
%
In recent years electric energy storage solutions gained a lot of attention due to the rising interest in renewable energy sources and the demand for mobile, electrical devices.
Supercapacitors with their high charging speed are a highly investigated candidate~\cite{shukla2000electrochemical, namisnyk2003survey, chee2016flexible}.
They comprise of porous electrodes, and an electrolyte consisting of electrically charged ions and a solvent.
When a potential difference is applied to the electrodes of a capacitor, the ions counteract the external influence by accumulating at the electrode of opposite charge.
There they form a layer opposing the surface charge on the electrodes, which, as a whole, is typically called electric double layer (\EDL{}).
These \EDL{}s have been studied for decades in numerous contexts and disciplines like physical interfaces in, biological membranes, or colloidal surfaces chemistry.
In recent times research focuses particularly on the \EDL{} structure which is of importance for charge storage in narrow confinements.

An important atomistic model to represent and study the structure of electrolytes is the primitive model (\PM{}).
In this model hard-core interactions between ions represent volume exclusion, and Coulomb interactions of point-charges within the hard cores capture the electrostatics.
The model often is studied by means of computer simulations in a controlled environment~\cite{valleau1980primitive, torrie1980electrical, fedorov2008ionic, kalcher2009structure, merlet2012molecular, merlet2012molecular}, but the thermodynamic insight gained from simulation studies is limited, compared to a rigorous theoretical approach.
A good candidate for such a theoretical framework that gives access to structure and thermodynamics of electrolytes is classical density-functional theory (\DFT{}) \cite{evans1979nature,hansen2013theory}.
Its key quantity is an energy functional of the one-body densities of the system that, applied to the equilibrium densities, yields the grand potential.
Even if in most systems the exact functional is not known, sophisticated approximate functionals were developed successfully for certain systems, in particular for rather simple particle interactions.
For instance, the 1D hard rod potential is one of the rare exactly solvable systems~\cite{percus1976equilibrium}, the hard-sphere potential is excellently treated in fundamental measure theory (\FMT{})~\cite{rosenfeld1989free, hansen2006density,roth2010fundamental}, the square well potential gives access to wetting phenomenons~\cite{van1989wetting}, and the Yukawa potential allows for the description of screened ions~\cite{hatlo2012density}.
For many interaction potentials, in particular short-ranged ones at low densities, mean-field approximations yield sufficiently accurate results, because the
potentials and correlations between particles decay rapidly at large interaction distances.
However, electrostatic interactions decay rather weakly and, at the same time, finite sized ions yield structure, especially close to electrodes and in dense systems. In consequence, the mean-field approach for the \PM{} produces rather poor quantitative predictions.

One way to avoid mean-field descriptions is utilizing the mean spherical approximation (\MSA{}), from which more accurate functionals for the \PM{} have been derived~\cite{waisman1972mean, hartel2015fundamental, yu2004density, roth2016shells, cats2021primitive,jiang2021revisitShellsOfCharge}.
The \MSA{} follows from a closure relation to the Ornstein-Zernike equation of liquid state theory \cite{hansen2013theory}.
The approximations made by such closure relations, however, are not as straight-forward as, for instance, approximations to the exact Barker-Henderson perturbation theory~\cite{barker1967perturbation}.
The electrostatic mean-field functional is the simplest term in this exact perturbation and follows from skipping all contributions of ``higher order''.
In this work, we go beyond the electrostatic mean-field approximation of the \PM{} by regarding these higher order terms in the Barker-Henderson perturbation instead of simply neglecting them. 
Such an approach has recently been applied to the hard-core Yukawa fluid, yielding very accurate density profiles, chemical potentials, and phase diagrams~\cite{tschopp2020mean}.

%
In the following, we start by introducing core concepts of \DFT{} and the \PM{} in \cref{sec:theory}.
We then derive the regular mean-field functional and our modified functionals from
the Barker-Henderson perturbation theory and give an outline on how to implement these functionals numerically.
In \cref{sec:results} we test different aspects of our modified functionals and compare our findings to those obtained from the regular mean-field functional, from
molecular dynamics simulation data, and from functionals that exploit the MSA closure. 
For our study we employ two geometries, one for an infinitely large parallel-plate capacitor and another one for the spherically symmetric surrounding of a fixed particle. 
The latter allows us to obtain the pair correlation function via the Percus trick instead of taking functional derivatives. 
Having at hand both routes to the correlation functions, we also test the consistency of the functionals by comparing
correlation functions obtained via the different routes.
As thermodynamic consistency checks we further examine the contact density theorem and the virial pressure formula. We conclude with a discussion of our results. 

\section{Theory}\label{sec:theory}
To model electrolytes we use the primitive model (\PM{}).
The solvent is modeled as dielectric background with relative permittivity $\epsilon$ which is accounted for in the electrostatic interactions of the ions via the Bjerrum length 
$\bjerrum = e^2/(4\pi\epsilon_0\epsilon\boltzmann T)$. The latter contains the elementary charge $e$, the vacuum permittivity $\epsilon_0$, Boltzmann's constant $\boltzmann$, and the temperature $T$. The pair potential of an ion $i$ of species $\specA$ with an ion $j$ of species $\specB$ at positions $\vec{r}_{\specA i}$ and $\vec{r}_{\specB j}$ consists of two parts: a purely repulsive hard-sphere interaction $v_{\specA\specB}^{\textrm{HS}}(r)$,
and the respective Coulomb potential $v_{\specA\specB}^{\textrm{ES}}(r)$ between point charges, sitting at the centers of the ions.
Accordingly, the resulting total pair potential reads
\begin{equation}\label{eqn:combinedpotential}
v_{\specA\specB}^{\textrm{PM}}(r) =
v_{\specA\specB}^{\textrm{HS}}(r) + v_{\specA\specB}^{\textrm{ES}}(r) =
\begin{cases}
\infty&r < \DADBTwo\\[.5em]
\frac{\bjerrum}{\beta}\frac{Z_{\specA}Z_{\specB}}{r}&r \geq \DADBTwo
\textrm{\,,}
\end{cases}
\end{equation}
where $r:=|\vec{r}_{\specA i}-\vec{r}_{\specB j}|$ is the distance between the two particles and $d_{\specA}$ are the hard-sphere diameters of the respective species $\specA$.
We further use the inverse temperature $\beta^{-1}=\boltzmann T$ as a thermal energy unit. 
Note that in \cref{eqn:combinedpotential} the Coulomb potential has no influence on the total pair potential at distances  \hbox{$r<\DADBTwo$}.
As we will see in the next section this is not the case for mean-field functionals in the framework of classical density-functional theory (\DFT{}) when the hard-sphere and electrostatic contributions are treated separately.

We are interested in two different geometries generated by certain external potentials. Each external potential consists of a hard impenetrable part and an electrostatic contribution, thus,

\begin{equation}
V_{\mathrm{ext},\specA}(\vec{r})
=
V_{\textrm{ext},\specA}^{\textrm{HS}}(\vec{r})
+
V_{\textrm{ext},\specA}^{\textrm{ES}}(\vec{r}) . 
\end{equation}
To model a parallel-plate capacitor (``$||$'') with potential difference $\Delta\Phi$ between the plates we employ hard walls in the $xy$-plane at $z=0$ and $z=L$ and a linear electrostatic potential within the walls, where the plates have the potentials $\pm\tfrac{\Delta\Phi}{2}$,
\begin{equation}\label{eqn:extpotwalls}
V_{\mathrm{ext},\specA}^{||}(z)=
\begin{cases}
\infty & \textrm{ for } z<\tfrac{d_{\specA}}{2} \textrm{ and } z>L-\tfrac{d_{\specA}}{2}\\ 
\frac{\Delta\Phi}{2}-\frac{\Delta\Phi}{L}z & \textrm{ else \,. }
\end{cases}
\end{equation}
Note that in \cref{sec:results} the electrostatic contribution is implemented as a boundary condition of the Poisson equation.
The other external potential describes a test-particle setup with a rotational symmetry (``$\circ$''), which leads to a spherical geometry.
The external potential fixes a particle with charge $Q_{\textrm{test}}$ at the origin and reads
\begin{equation}\label{eqn:extpotpercus}
V_{\mathrm{ext},\specA}^{\circ}(r)=
\begin{cases}
\infty & \textrm{ for } r<(d_{\textrm{test}} + d_{\specA})/2\\ 
\bjerrum Z_{\specA}\frac{Q_{\textrm{test}}}{r} & \textrm{ else \,.}
\end{cases}
\end{equation}
In this geometry the external potential is implemented directly in \cref{sec:results} and only the remaining boundary conditions for the Poisson equations are calculated in \cref{app:numerical}.

Theoretically, we describe electrolyte systems in the thermodynamic grand canonical ensemble.
It allows for heat and particle exchange, which means that the thermodynamic potential $\Omega(T, V, \mu)$ depends on the temperature $T$, the system volume $V$, and the chemical potential $\mu$. In a system with charged electrodes an additional pair of conjugated variables that contribute to the grand potential is given by the electrode charge and the electrode potential~\cite{van2012statistical}.

The key quantity of \DFT{} is the one-body density distribution
\begin{equation}
\rho_{\specA}^{(1)}(\vec{r})
=
\left\langle\sum^{N_{\specA}}_{i=1}\delta(\vec{r}-\vec{r}_{\specA{}i})\right\rangle
\textrm{\,,}
\end{equation}
which is defined as an ensemble average (or classical trace, denoted by $\langle\dots\rangle$) over the density operator, a sum of $\delta$-distributions over all particle positions $\vec{r}_{\specA{}i}$ \cite{hansen2013theory_delta}.
Similarly the $n$-body density distribution $\rho^{(n)}(\vec{r}_1,\dots,\vec{r}_n)$ is obtained. 
The two-body density distribution is related to the pair-correlation function 
\begin{equation}
g_{\specA\specB}^{(2)}(\vec{r}^{\,},\vec{r}^{\,\prime})
=
\frac{
\rho_{\specA\specB}^{(2)}(\vec{r}^{\,},\vec{r}^{\,\prime})
}{
\rho_{\specA}^{(1)}(\vec{r}^{\,})\rho_{\specB}^{(1)}(\vec{r}^{\,\prime})
}
\textrm{\,,}
\end{equation}
which shows the local density fluctuations of a fluid compared to an ideal gas. 
It is directly related to the total pair-correlation function $h^{(2)}$ by $h^{(2)}=g^{(2)}-1$. For convenience we use the notation $\rho_{\specA}:=\rho^{(1)}_{\specA}$ in the remaining work. 
Further, direct correlation functions $c^{(2)}$ are defined via the Ornstein-Zernike (\OZ{}) equation~\cite{ornstein1914accidental}
\begin{equation} \label{eqn:OZ}
\begin{split}
h_{\specA\specB}^{(2)}(\vec{r},\vec{r}^{\,\prime})
&=
c_{\specA\specB}^{(2)}(\vec{r},\vec{r}^{\,\prime})\\
&+
\sum_{\specD}\int_{\mathbb{R}^3}\inftes{\vec{r}^{\,\prime\prime}}
c_{\specA\specD}^{(2)}(\vec{r},\vec{r}^{\,\prime\prime})\rho_{\specD}(\vec{r}^{\,\prime\prime})
h_{\specD\specB}^{(2)}(\vec{r}^{\,\prime\prime},\vec{r}^{\,\prime})
\textrm{\,.}
\end{split}
\end{equation}
These direct correlation functions emerge from the direct interaction of two closely situated particles. Together with the indirect correlations via one or more intermediate particles, as covered by the second term on the right-hand side of \cref{eqn:OZ}, they contribute to the total correlation function. 

Since, in general, the pair-correlation function is obtained from the one- and two-body density distributions, Percus showed that it can also be obtained without knowledge of $\rho^{(2)}$ by a simple trick~\cite{percus1962approximation, frisch1964equilibrium}.
A (fixed) external potential $v_{\textrm{ext},\specA}$, that resembles the pair interaction of one particle of species $\specA$ with the surrounding particles of species $\specB$, is placed at the origin of the system.
Then the one-particle density distribution of the particles in the system normalized with respect to the respective bulk density 
of the homogeneous system yields the pair correlation function, 
\begin{equation}\label{eqn:percus}
g_{\specA\specB}^{(2)}(\vec{r}, \vec{r}^{\,\prime})
=
\frac{
\rho_{\specB}^{(1)}(\vec{r}^{\,\prime}|v_{\textrm{ext},\specA})
}{
\rho_{\specB}^{(1)}(\vec{r}^{\,}|v_{\textrm{ext},\specA}=0)
}
\textrm{\,.}
\end{equation}
In other words, the one-body distribution of the particles surrounding the ``fixed particle'' in the origin, represented by the external field, yields the required two-body distribution.
%
\subsection{Classical Density-Functional Theory}\label{sec:theory:cdft}
%
In the framework of \DFT{} one minimizes an energy functional $\Omega\funcdepend$ of the local particle densities $\rho_{\specA}(\vec{r})$.
In its minimum this functional equals the grand potential $\Omega$ and the densities that minimize the functional are the equilibrium densities~\cite{hohenberg1964inhomogeneous, mermin1965thermal, evans1979nature}.
The energy functional can be written in the form
\begin{equation}\label{eqn:grandpotential}
\begin{split}
\Omega\funcdepend =&
\sum_{\specA}\int\inftes{\vec{r}}\rho_{\specA}(\vec{r})
(V_{\textrm{ext},\specA}(\vec{r}) - \mu_{\specA})\\
&+\mathcal{F}_{\textrm{id}}\funcdepend
+\mathcal{F}_{\textrm{ex}}\funcdepend\textrm{\,.}\\
\end{split}
\end{equation}
The first term contains the external potential $V_{\textrm{ext}}$
that couples to every particle individually and
the species-dependent chemical potential $\mu_{\specA}$.
$\mathcal{F}_{\textrm{id}}$ captures all contributions except those arising from particle interactions and approaches the ideal (gas) free energy in the minimum of the functional.
This ideal free-energy functional is given by
\begin{equation}\label{eqn:idealenergy}
\beta\mathcal{F}_{\textrm{id}}\funcdepend =
\sum_{\specA}\int\inftes{\vec{r}}\rho_{\specA}(\vec{r})
\left(\ln(\rho_{\specA}(\vec{r})\Lambda_{\specA}^3) -1\right)
\textrm{\,,}
\end{equation}
where $\Lambda_{\specA}=h/\sqrt{2\pi m_{\specA}\boltzmann T}$ is the thermal wavelength, $h$ is Planck's constant, and $m_{\specA}$ is the mass of a particle of species $\specA$. All particle interactions are taken into account in the last term of \cref{eqn:grandpotential}, which is usually called \emph{excess term} and covers all contributions that excess those of the ideal gas. Its second functional derivative
gives the aforementioned direct correlation functions,
\begin{equation}\label{eqn:directcorrelation}
c_{\specA\specB}^{(2)}(\vec{r}^{\,}, \vec{r}^{\,\prime}) =
-\beta\frac{\delta^{2}\mathcal{F}_{\textrm{ex}}\funcdepend}
{\delta\rho_{\specA}(\vec{r}^{\,})\delta\rho_{\specB}(\vec{r}^{\,\prime})}\textrm{\,.}
\end{equation}
Thus, in addition to the Percus trick, we can obtain the total correlation function from DFT via the direct correlation functions and solving the OZ equation.
We mention that higher-order functional derivatives of of the excess term define a whole hierarchy of $n$-body direct correlation functions. 
%
\subsection{Excess Functionals}\label{sec:theory:excess}
%
The excess functional $\mathcal{F}_{\textrm{ex}}$ of \cref{eqn:grandpotential} gives the excess free energy in the minimum of the functional.
In the case of the hard-sphere model several versions of sophisticated functionals exist, for instance \cite{tarazona2000density, hansen2006density}, which are typically based on fundamental measure theory \cite{rosenfeld1989free, roth2010fundamental}.
In our numerical analysis in \cref{sec:results} we use the \emph{White Bear mark II} functional~\cite{hansen2006density} with an correction by Tarazona~\cite{tarazona2000density}, because the resulting functional fulfills thermodynamic sum rules and predicts thermodynamic properties like the bulk pressure, the liquid-solid phase transition, pair-correlation structures in the vicinity of walls, and the interfacial free energy very accurately~\cite{hansen2006density,oettel2010free, hartel2012tension,hartel2015anisotropic}.
\subsubsection{Barker-Henderson Perturbation Theory}
\label{sec:theory:excess:perturbation}
For the \PM{}, we need to construct a functional that takes into account both the hard-sphere and the Coulomb interactions. Since a hard-sphere system can be represented very accurately by the aforementioned functional, we treat it as a reference system. Then the Coulomb interactions are added as a perturbation. Accordingly, the decomposition of the perturbed pair potential has the form 
\begin{equation}\label{eqn:perturbationpotential}
v_{\specA\specB}(\vec{r},\vec{r}^{\,\prime};\lambda)=
v_{\specA\specB}^{\textrm{HS}}(\vec{r},\vec{r}^{\,\prime})+
\lambda v_{\specA\specB}^{\textrm{pert}}(\vec{r},\vec{r}^{\,\prime})
\textrm{\,,}\quad
0\leq\lambda\leq1
\textrm{\,,}
\end{equation}
where $\lambda$ is a perturbation measure that ``switches on'' the perturbation.
Note, that due to symmetry we can use $v_{\specA\specB}(\vec{r},\vec{r}^{\,\prime}):=v_{\specA\specB}(|\vec{r}-\vec{r}^{\,\prime}|)$, without loss of generality. 
Now, the corresponding excess free energy functional is given by
\begin{equation}\label{eqn:perturbation}
\begin{split}
\mathcal{F}_{\textrm{ex}}& =
\mathcal{F}_{\textrm{ex}}^{\textrm{HS}}
+\mathcal{F}_{\textrm{ex}}^{\textrm{pert}}
+\mathcal{F}_{\textrm{ex}}^{\textrm{corr}}
\textrm{\,,}
\end{split}
\end{equation}
which can be derived from the Barker-Henderson perturbation theory~\cite{barker1967perturbation,hansen2013theory_perturbation} and also follows from a functional integration of $\delta\mathcal{F}^{\text{ex}}/\delta v_{\specA\specB}(\vec{r},\vec{r}^{\prime})$ over $\lambda$~\cite{evans1979nature}. The involved functionals read
\begin{equation}\label{eqn:perturbationterm}
\begin{split}
\mathcal{F}_{\textrm{ex}}^{\textrm{pert}}\funcdepend
=&\frac{1}{2}\sum_{\specA\specB}
\iint\!\inftes{\vec{r}_1}\inftes{\vec{r}_2}
\rho_{\specA}(\vec{r}_1)\rho_{\specB}(\vec{r}_2)\\
&\times
v_{\specA\specB}^{\textrm{pert}}(|\vec{r}_1-\vec{r}_2|)
\end{split}
\end{equation}
and
\begin{equation}
\begin{split}
\mathcal{F}_{\textrm{ex}}^{\textrm{corr}}\funcdepend
=&
\frac{1}{2}\sum_{\specA\specB}\int_0^1\!\inftes{\lambda}
\iint\!\inftes{\vec{r}_1}\inftes{\vec{r}_2}
\rho_{\specA}(\vec{r}_1)\rho_{\specB}(\vec{r}_2) \\
&\times
v_{\specA\specB}^{\textrm{pert}}(|\vec{r}_1-\vec{r}_2|)
h^{(2)}_{\specA\specB}(\vec{r}_1,\vec{r}_2; \lambda)
\textrm{\,.} \label{eqn:correlationterm}
\end{split}
\end{equation}
The first term on the right-hand side of \cref{eqn:perturbation} contains the hard-sphere excess free-energy functional and the second term contains a mean-field treatment of the perturbation potential, as given in \cref{eqn:perturbationterm}.
The third term, as given in \cref{eqn:correlationterm}, occurs due to the non-linear behavior of the excess free energy when adding two pair potentials and is called correlation term. 
Here, $h^{(2)}_{\specA\specB}(\vec{r}_1,\vec{r}_2;\lambda)$ is the total pair-correlation function in a system with the equilibrium density that minimizes the functional at full perturbation but with pair potential
$v_{\specA\specB}(\vec{r};\lambda)$ depending on $\lambda$.
All free energy contributions that arise from correlations between the two potentials $v_{\specA\specB}^{\textrm{HS}}$ and $v_{\specA\specB}^{\textrm{pert}}$ are treated within \cref{eqn:correlationterm}.

\begin{figure}
\centering
\includegraphics[width=\linewidth]{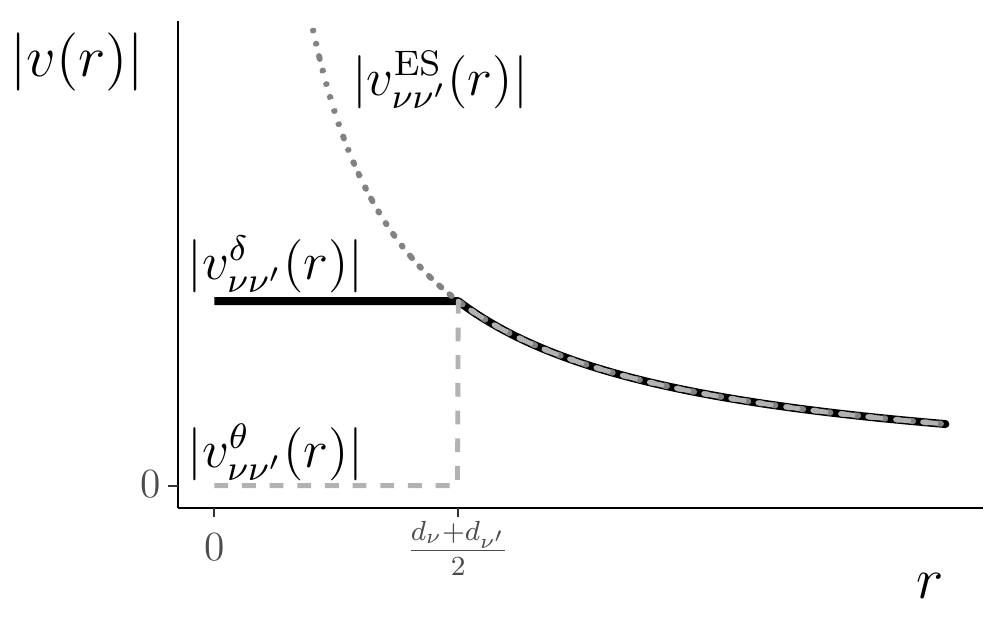}
\caption{Sketch for the absolute value of different electrostatic pair potentials $v_{\specA\specB}$ that all yield the pair potential $v_{\specA\specB}^{\textrm{PM}}$ of the \PM{}, when combined with pair potentials of hard spheres with diameters $d_{\specA}$, given in \cref{eqn:combinedpotential}.}\label{fig:potentials}
\end{figure}

We stress that the decomposition in \cref{eqn:perturbationpotential} is not unique for a pair potential as given in \cref{eqn:combinedpotential} for a reference hard-sphere system. Obviously, arbitrary functions that vanish on $[(\DADB)/2,\infty)$ can be added to the interaction potential $v_{\specA\specB}(r)$ without changing its value that is infinite for $r<(\DADB)/2$.
Some examples for perturbation potentials are shown in \cref{fig:potentials}.
While all these different functions lead to the same pair potential $v_{\specA\specB}^{\textrm{PM}}$ when added to the hard-sphere potential $v_{\specA\specB}^{\textrm{HS}}$, they result in different perturbation functionals $\mathcal{F}_{\textrm{ex}}^{\textrm{pert}}$ and $\mathcal{F}_{\textrm{ex}}^{\textrm{corr}}$.

\subsubsection{Mean-Field Electrostatics Functional}
\label{sec:theory:excess:meanfield}
As a simple approximation of \cref{eqn:perturbation} one can neglect the correlation term $\mathcal{F}_{\textrm{ex}}^{\textrm{corr}}$.
If we are interested in a \PM{} with pair-interaction potential $v_{\specA\specB}^{\textrm{PM}}$ as given in \cref{eqn:combinedpotential} by $v_{\specA\specB}^{\textrm{HS}}+v_{\specA\specB}^{\textrm{ES}}$, the respective free-energy functional becomes $\mathcal{F}_{\textrm{ex}}=\mathcal{F}_{\textrm{ex}}^{\textrm{HS}}+\mathcal{F}_{\textrm{ex}}^{\textrm{MF}}$ with $\mathcal{F}_{\textrm{ex}}^{\textrm{pert}}$ being the mean-field Coulomb free-energy functional 
\begin{equation}\label{eqn:f-mf}
\mathcal{F}_{\textrm{ex}}^{\textrm{MF}}\funcdepend=
\frac{\bjerrum}{2\beta}\sum_{\specA\specB}Z_{\specA}Z_{\specB}
\iint\!\inftes{\vec{r}_1}\inftes{\vec{r}_2}
\frac{\rho_{\specA}(\vec{r}_1)\rho_{\specB}(\vec{r}_2)}
{|\vec{r}_1-\vec{r}_2|}\\
\textrm{\,.}
\end{equation}
This approach is commonly used and, for vanishing hard-sphere contributions, results in a functional that is minimized by the density profiles that solve Poisson-Boltzmann equation. (This can be seen by rewriting the functional in terms of the electrostatic potential and plugging it into the Euler equations shown in \cref{eqn:update}~\cite{hansen2013theory_pb}.)
The Coulomb mean-field functional from eq.~(\ref{eqn:f-mf}) can be used as a starting point from which more sophisticated functionals can be created by adding non-vanishing correlation terms.
\subsubsection{Core-Corrected Functional}
\label{sec:theory:excess:theta}
To take advantage of the correlation term in \cref{eqn:perturbation}, one can 
establish a low-density treatment of \cref{eqn:correlationterm} by applying 
\begin{equation}\label{eqn:lowdensity}
\int_0^1\!\inftes{\lambda} h^{(2)}_{\specA\specB}(\vec{r}_1,\vec{r}_2; \lambda)
\approx
-\theta\left(\DADBTwo - |\vec{r}_1-\vec{r}_2|\right)\textrm{\,,}
\end{equation}
where the total correlation function does not depend on the switching parameter $\lambda$ and has the form of a low-density hard-sphere fluid. The involved Heaviside step function $\theta$ is defined by \hbox{$\theta(r)=1$} for \hbox{$r\geq 0$} and \hbox{$\theta(r)=0$} for \hbox{$r<0$}.
If we use this approximation and consider again the pair potential of the \PM{} from \cref{eqn:combinedpotential}, then the correlation term $\mathcal{F}_{\textrm{ex}}^{\textrm{corr}}$ has a similar form as the mean-field term from \cref{eqn:f-mf} and merging both terms yields the core-corrected mean-field functional 
\begin{equation}\label{eqn:theta}
\begin{split}
\mathcal{F}_{\textrm{ex}}^{\theta}\funcdepend=&
\frac{\bjerrum}{2\beta}\sum_{\specA\specB}Z_{\specA}Z_{\specB}
\iint\inftes{\vec{r}_1}\inftes{\vec{r}_2}
\frac{\rho_{\specA}(\vec{r}_1)\rho_{\specB}(\vec{r}_2)}
{|\vec{r}_1-\vec{r}_2|}\\
&\times \theta\left(|\vec{r}_1-\vec{r}_2| - \DADBTwo\right)
\end{split}
\end{equation}
with \hbox{$\mathcal{F}_{\textrm{ex}}=\mathcal{F}_{\textrm{ex}}^{\textrm{HS}}+\mathcal{F}_{\textrm{ex}}^{\theta}$}.
Due to the infinite repulsiveness of the hard spheres, forbidden areas in phase space exist that cannot be visited by the system compared to a system without hard-sphere repulsion.
Thus, one can interpret the result in \cref{eqn:theta} as a phase-space restriction of the pure mean-field term, where electrostatic mean-field contributions from the forbidden areas in phase space are ignored.

Interestingly, the functional in \cref{eqn:theta} can also be obtained if the perturbation potential
\begin{equation}\label{eqn:espotentialtheta}
v_{\specA\specB}^{\theta}(r) = 
\begin{cases}
0 & r < \DADBTwo \\
v_{\specA\specB}^{\textrm{ES}}(r) & r \geq \DADBTwo
\end{cases}
\end{equation}
is used in \cref{eqn:perturbationpotential}. This perturbation now is a Coulomb potential with a cut out center and leads to a functional where the perturbation term $\mathcal{F}_{\textrm{ex}}^{\textrm{pert}}$ already has the form of $\mathcal{F}_{\textrm{ex}}^{\theta}$, as given in \cref{eqn:theta}. The correlation term can simply be ignored. 

Note that we can further get rid of the Heaviside function in \cref{eqn:theta} by modifying the integration volume for $\vec{r}_2$ in this equation from the entire space $\mathbb{R}^3$ to 
\begin{equation}\label{eqn:region}
\mathcal{V}^{\prime}:=\mathcal{V}^{\prime}(\vec{r}_{1})=\left\{\vec{r}_2\in\mathbb{R}^3\big|
|\vec{r}_1-\vec{r}_2|\geq \DADBTwo\right\}
\textrm{ ,}
\end{equation}
which is \emph{the entire space without the volume of a sphere with radius $\DADBTwo$ and center at $\vec{r}_1$}. Accordingly, the functional now reads 
\begin{equation}\label{eqn:theta2}
\mathcal{F}_{\textrm{ex}}^{\theta}\funcdepend=
\frac{\bjerrum}{2\beta}\sum_{\specA\specB}Z_{\specA}Z_{\specB}\!
\int\!\inftes{\vec{r}_1}\int_{\mathcal{V}^{\prime}}\!\inftes{\vec{r}_2}
\frac{\rho_{\specA}(\vec{r}_1)\rho_{\specB}(\vec{r}_2)}
{|\vec{r}_1-\vec{r}_2|}
\textrm{\,.}
\end{equation}
While the functional $\mathcal{F}_{\textrm{ex}}^{\theta}$ has successfully been utilized in the past to study under-screening in dense electrolytes via its second functional derivative and the OZ relation~\cite{coupette2018screening}, it has some significant downsides that we will elaborate in \cref{sec:results}. 
We will see that this functional has low predictive capabilities in almost all situations we studied. In particular, it overestimates contact values as well as layering effects in the vicinity of hard walls greatly.
\subsubsection{Continuous Core Correction}
\label{sec:theory:excess:delta}

In the previous section we used the discontinuous perturbation potential $v_{\specA\specB}^{\theta}$ for the electrostatic contribution to the pair potential $v_{\specA\specB}^{\textrm{PM}}$, where the potential vanishes when the hard cores overlap. Instead, we now follow the approach by Weeks~\etal~\cite{weeks1971role}, who split pair potentials in a way such that the soft perturbation is continuous within the hard core.
We apply this idea to $v_{\specA\specB}^{\textrm{PM}}$ by splitting of the continuous electrostatic pair potential
\begin{equation}\label{eqn:espotentialdelta}
v_{\specA\specB}^{\delta}(r) = 
\begin{cases}
v_{\specA\specB}^{\textrm{ES}}\left(\DADBTwo\right)& r < \DADBTwo \\
v_{\specA\specB}^{\textrm{ES}}(r) & r \geq \DADBTwo
\textrm{\,,}
\end{cases}
\end{equation}
which gives the constant value of the Coulomb pair potential at particle contact for two particles in the core region, as shown in \cref{fig:potentials}.
The identical potential can be achieved by a point-like test charge in the vicinity of a homogeneously charged spherical shell of radius $(\DADB)/2$.
If the test charge resides inside the shell, the potential is constant.
On the outside the test particle experiences the regular Coulomb potential of the entire charge of the shell.
A similar ansatz was used to modify the Poisson-Boltzmann theory for point charges such that it respects correlations from steric interactions~\cite{forsman2004cpb,de2020interfacial}.
The resulting perturbation functional is 
\begin{equation}\label{eqn:delta}
\begin{split}
\mathcal{F}_{\textrm{ex}}^{\delta}\funcdepend
&=
\mathcal{F}_{\textrm{ex}}^{\theta}\funcdepend\\
&+
\frac{\bjerrum}{\beta}\sum_{\specA\specB}
\frac{Z_{\specA}Z_{\specB}}{\DADB}
\int\!\inftes{\vec{r}_1}\!\!
\int_{\mathbb{R}^3\setminus\mathcal{V}^{\prime}}\!\!\!\inftes{\vec{r}_2}
\rho_{\specA}(\vec{r}_1)\rho_{\specB}(\vec{r}_2)
\textrm{\,.}
\end{split}
\end{equation}

As it turns out, the functional ${\mathcal{F}}_{\textrm{ex}}^{\delta}$ performs much better than the mean-field functional ${\mathcal{F}}_{\textrm{ex}}^{\textrm{MF}}$. In \cref{sec:results} we will see that it, for example, predicts the structure of electric double layers well for strong electrostatic interactions and high densities. This difference in performance originates from the different pair potentials $v_{\nu\nu'}^{\theta}$ and $v_{\nu\nu'}^{\delta}$, as defined in \cref{eqn:espotentialtheta,eqn:espotentialdelta}. 
While the latter is continuous, the former has a discontinuity at the contact separation that leads to the poor performance.

We remind that the potential in \cref{eqn:espotentialdelta} as well as all potentials shown in \cref{fig:potentials} result in the same Hamiltonian, respectively, and, thus, in the same physics. However, the corresponding functionals $\mathcal{F}_{\textrm{ex}}^{\textrm{pert}}$ and $\mathcal{F}_{\textrm{ex}}^{\textrm{corr}}$ differ and, consequently, their approximations can predict contradictory behavior for one and the same system. 

\subsection{Functional Derivatives}\label{sec:theory:funcderivs}

In DFT, the variational principle applied to the grand potential energy functional $\Omega\funcdepend$,
\begin{equation}
\left.
\frac{\delta\Omega\funcdepend}{\delta\rho_{\specA}(\vec{r})}
\right|_{\{\rho_{\specC}\}=\{\mbox{\tiny equilibrium densities}\}}=0
\textrm{\,,}
\end{equation}
yields Euler-Lagrange equations. 
By defining a modified chemical potential \hbox{$\mu^{*}_{\specA} = \mu_{\specA}-\boltzmann T\ln(\Lambda_{\specA}^3\rho_{\specA})$} for chemical potentials $\mu_{\specA}$ and corresponding bulk densities $\rho_{\specA}$ of respective species $\specA$, the Euler-Lagrange equations for the equilibrium densities read 
\begin{equation}\label{eqn:update}
\rho_{\specA}(\vec{r}) = \rho_{\specA}\exp(
\beta\mu_{\specA}^{*} - \beta V_{\textrm{ext}, \specA}(\vec{r}) -
\beta\fdv{\mathcal{F}_{\textrm{ex}}\funcdepend}{\rho_{\specA}(\vec{r})})
\textrm{\,.}
\end{equation}
These equations are update equations which are used to obtain numerical solutions by means of Picard-iterations.
Here, the calculation of the right hand side of \cref{eqn:update} requires the computation of the first functional derivative of the excess free energy functional. In the following we present these functional derivatives for the previously discussed functionals $\mathcal{F}_{\textrm{ex}}^{\textrm{MF}}$, 
$\mathcal{F}_{\textrm{ex}}^{\theta}$, and 
$\mathcal{F}_{\textrm{ex}}^{\delta}$; for details on the functional derivative of the hard-sphere term 
$\mathcal{F}_{\textrm{ex}}^{\textrm{HS}}$ we refer to previous work \cite{hartel2013density}.

The functional derivative of the mean-field functional from \cref{eqn:f-mf} is 
\begin{equation}\label{eqn:mf-deriv}
\fdv{{\mathcal{F}}_{\textrm{ex}}^{\textrm{MF}}\funcdepend}{\rho_{\specA}(\vec{r})} = 
\frac{\bjerrum Z_{\specA}}{\beta}\sum_{\specB}
\int\!\inftes{\vec{r}^{\,\prime}}
\frac{Z_{\specB}\rho_{\specB}(\vec{r}^{\,\prime}))}
{|\vec{r}-\vec{r}^{\,\prime}|}
\textrm{\,.}
\end{equation}
The right-hand side (\rhs{}) of \cref{eqn:mf-deriv} is related to the dimensionless electrostatic potential $\Phi(\vec{r})$ of the system which is related to the charge distribution in the system via the Poisson equation 
\begin{equation}\label{eqn:poisson}
\laplacian\Phi(\vec{r}) = -4\pi\bjerrum\sum_{\specA}Z_{\specA}\rho_{\specA}(\vec{r})
\end{equation}
and respective electrostatic boundary conditions of the system
(e.g. surface charge density of a plate capacitor). 
The formal solution to this equation can be obtained by means of  Green's functions, which simplifies the above functional derivative to 
\begin{equation}\label{eqn:meanfieldderiv}
\fdv{{\mathcal{F}}_{\textrm{ex}}^{\textrm{MF}}\funcdepend}{\rho_{\specA}(\vec{r})} = 
\frac{Z_{\specA}}{\beta}\Phi(\vec{r})
\end{equation}
by replacing the integral with the formal solution. 

\begin{figure}
\hspace{-.05\textwidth}
\includegraphics[width=.35\textwidth]{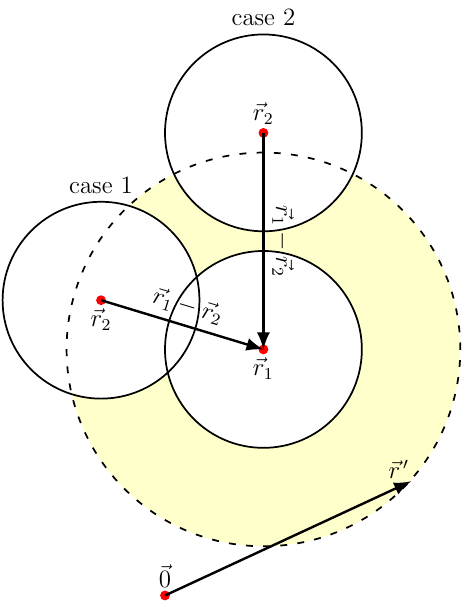}
\caption{Sketch to depict the relevant cases in the Shell Theorem, \cref{eqn:shelltheorem}.
Integration on the left-hand side (\lhs{}) is over the dashed shell. In case 1, the charge located in $\vec{r}_2$ is inside the shell, which results in a constant.
In case 2, the charge is located outside the shell and, thus, simply feels the field of a total charge located in $\vec{r}_1$.}
\label{fig:shelltheorem}
\end{figure}

In order to obtain the functional derivative of our modified mean-field functionals $\mathcal{F}_{\textrm{ex}}^{\theta}$ and $\mathcal{F}_{\textrm{ex}}^{\delta}$, we first rewrite \cref{eqn:theta2} in a form that simplifies the derivations. 
For this purpose, we apply the Shell Theorem\phantom{\tiny\cite{tipler2007physics}\!\!\!\!\!\!\!\!\!\!}
\footnote{To see that this holds true one can substitute $\vec{r}:=\vec{r}^{\,\prime}-\vec{r}_1$ and $\vec{r}_{12}:=\vec{r}_2-\vec{r}_1$. Then the formal solution to a spherical charge shell is obtained. The solution to this problem is, for example, given in~\cite{tipler2007physics}.},
\begin{equation}\label{eqn:shelltheorem}
\begin{split}
&\frac{1}{\pi(\DADB)^2}\int\inftes{\vec{r}^{\,\prime}}
\frac{\delta\left(|\vec{r}^{\,\prime}-\vec{r}_1|-\DADBTwo\right)}
{|\vec{r}_2-\vec{r}^{\,\prime}|}\\
&\\
&=
\begin{cases}
\frac{2}{\DADB} & |\vec{r}_1-\vec{r}_2|<\DADBTwo\\
&\\
\frac{1}{|\vec{r}_1-\vec{r}_2|} & |\vec{r}_1-\vec{r}_2|\geq\DADBTwo
\textrm{\,,}
\end{cases}
\end{split}
\end{equation}
which is depicted in \cref{fig:shelltheorem}.
We replace the term $\tfrac{1}{|\vec{r}_1-\vec{r}_2|}$ in the functional in \cref{eqn:theta2} by the left-hand side (\lhs{}) of \cref{eqn:shelltheorem}. Note that this substitution is possible because the integration in \cref{eqn:theta2} is restricted to the volume $\mathcal{V}^{\,\prime}$.

To compute the functional derivatives we further aim to remove the restriction to the volume $\mathcal{V}^{\,\prime}$ from \cref{eqn:theta2} and complete the integral to cover the entire space instead of the restriction to $\mathcal{V}^{\,\prime}$.
For this purpose, we add a zero to the \rhs{} integral of \cref{eqn:theta2} by extending the integration volume over $\vec{r}_2$ to the entire space and subtracting the same integral with an integration volume $\mathbb{R}^3\setminus\mathcal{V}^{\,\prime}$ for $\vec{r}_2$ (which finally leads to the last term of \cref{eqn:f-theta}). In this subtracted term we apply the shell theorem again, but now only the first case on the e of \cref{eqn:shelltheorem} applies.
As a result we find
\begin{equation}\label{eqn:f-theta}
\begin{split}
\functionaltheta\funcdepend=&
\frac{\bjerrum}{2\beta}\sum_{\specA\specB}Z_{\specA}Z_{\specB}
\bigg(\int\inftes{\vec{r}_1}\int\inftes{\vec{r}_2}\int\inftes{\vec{r}^{\,\prime}}\\
\times&\frac{\rho_{\specA}(\vec{r}_1)\rho_{\specB}(\vec{r}_2)
\delta\left(|\vec{r}^{\,\prime}-\vec{r}_1|-\DADBTwo\right)}
{\pi(\DADB)^2|\vec{r}_2-\vec{r}^{\prime}|}\\
-&2\int\inftes{\vec{r}_1}\!\int_{\mathbb{R}^3\setminus\mathcal{V}^{\,\prime}}\!\inftes{\vec{r}_2}
\frac{\rho_{\specA}(\vec{r}_1)\rho_{\specB}(\vec{r}_2)}{\DADB}\bigg)
\textrm{\,,}
\end{split}
\end{equation}
where the last term on the right-hand-side is the discussed subtracted term. 

Now, we can take the functional derivative of $\functionaltheta$
and find
\begin{equation}\label{eqn:thetaderiv}
\begin{split}
\fdv{\functionaltheta\funcdepend}{\rho_{\specA}(\vec{r})}&=
\frac{Z_{\specA}}{2\beta}\sum_{\specB}
\left(\phi_{\specB}\ast\tilde{\delta}_{\specA\specB}\right)(\vec{r})\\
&+\frac{Z_{\specA}}{2\beta}\sum_{\specB}\bjerrum\int\inftes{\vec{r}^{\,\prime}}
\frac{n_{\specA\specB}^{\delta}(\vec{r}^{\,\prime})}{|\vec{r}-\vec{r}^{\,\prime}|}\\
&-\frac{\bjerrum Z_{\specA}}{\beta}\sum_{\specB}n_{\specA\specB}^{\theta}(\vec{r})
\textrm{\,,}
\end{split}
\end{equation}
where we defined the weighted densities
\begin{subequations}\label{eqn:weighteddensities}
\begin{equation}
n_{\specA\specB}^{\delta}(\vec{r})=
\left(Z_{\specB}\rho_{\specB}\ast\tilde{\delta}_{\specA\specB}\right)(\vec{r})
\end{equation}
and
\begin{equation}
n_{\specA\specB}^{\theta}(\vec{r})=
\left(Z_{\specB}\rho_{\specB}\ast\tilde{\theta}_{\specA\specB}\right)(\vec{r}) \textrm{\,.}
\end{equation}
\end{subequations}
Here, $f\ast g$ denotes a convolution of functions $f$ and $g$, i.e. $(f\ast g)(\vec{r})=\int \textrm{d}\vec{r}'f(\vec{r}')g(\vec{r}-\vec{r}')$, and $\tilde{\theta}_{\specA\specB}$ and $\tilde{\delta}_{\specA\specB}$ are spherical Heaviside functions \hbox{$\tilde{\theta}_{\specA\specB}(\vec{r}):=\tfrac{2}{\DADB}\,\theta(\DADBTwo - |\vec{r}|)$} and \hbox{$\delta$-distributions} \hbox{$\tilde{\delta}_{\specA\specB}(\vec{r}):=\tfrac{1}{\pi(\DADB)^2}\delta(|\vec{r}|-\DADBTwo)$}.
Thus, the first term on the \rhs{} of \cref{eqn:thetaderiv} contains the convolution
of a spherical \hbox{$\delta$-distribution} with 
a species-specific electrostatic potential 
\begin{equation}\label{eqn:speciespotential}
\phi_{\specA}(\vec{r})
=\bjerrum \int\inftes{\vec{r}^{\,\prime}}
\frac{Z_{\specA}\rho_{\specA}(\vec{r}^{\,\prime})}{|\vec{r}-\vec{r}^{\,\prime}|}
\end{equation}
that only takes the charge density of the respective particle species into account; of course, both charge-specific quantities are connected via Poisson's equation and all species-specific electrostatic potentials add up to the total electrostatic potential $\Phi$. 
Similarly, the second term on the \rhs{} of \cref{eqn:thetaderiv} contains potentials of the weighted densities $n_{\specA\specB}^{\delta}$ and, finally, the last term consists of weighted densities $n_{\specA\specB}^{\theta}$. 

The careful reader might worry about the numerical feasibility of \cref{eqn:speciespotential}, because bulk charge densities of a single species would produce diverging potential values. This, however, can be circumvented by decomposing the charge densities into a bulk part $\rho_{\specA}$ and an \emph{excess} part $\Delta\rho_{\specA}(\vec{r})$,
\begin{equation}
\rho_{\specA}(\vec{r}) = \rho_{\specA} + \Delta\rho_{\specA}(\vec{r})
\textrm{\,.}
\end{equation}
As a consequence, bulk contributions to the different species-specific potentials cancel each other due to the charge neutrality of bulk.
At the same time, the structure of the functional derivatives remains unchanged, just all occurrences of $\rho_{\specA}(\vec{r})$ straightforwardly must be replaced by $\Delta\rho_{\specA}(\vec{r})$.
In addition, the bulk part of the densities leads to a separate term that equals the reduced chemical potential and, accordingly, cancels the identical term $\beta\mu_{\specA}^{\ast}$ that occurs in~\cref{eqn:update}.
For the $\functionaltheta$ functional the reduced chemical potentials become
\begin{equation}
\label{eqn:mu-theta}
\begin{split}
\mu_{\specA}^{\ast\theta}&=
\left.\fdv{\functionaltheta\funcdepend}{\rho_{\specA}(\vec{r})}
\right|_{\rho_{\specA}(\vec{r})=\rho_{\specA}}\\
&=\frac{\pi}{6}\frac{Z_{\specA}\bjerrum}{\beta}
\sum_{\specB}Z_{\specB}\rho_{\specB}(\DADB)^2
\textrm{\,,}
\end{split}
\end{equation}
as we show in detail in \cref{app:thetaderiv}.
The $\mu_{\specA}^{\ast}$ vanish as long as charge neutrality is given and the hard-sphere diameters of the species are identical.

Now, in order to derive similar results for the functional ${\mathcal{F}}_{\textrm{ex}}^{\delta}$, we use the continuous pair potential $v_{\specA\specB}^{\delta}$ from \cref{eqn:espotentialdelta} with non-vanishing core contribution instead of $v_{\specA\specB}^{\theta}$.
Consequently, an additional contribution appears in the perturbation functional $\mathcal{F}_{\textrm{ex}}^{\textrm{pert}}$, which happens to be equal to the last term from \cref{eqn:f-theta}, but with opposing sign. This leads to a cancellation of these terms leaving only the first term of \cref{eqn:f-theta}.
As a result, we find
\begin{equation}\label{eqn:f-delta}
\begin{split}
\functionaldelta\funcdepend=&
\frac{\bjerrum}{2\beta}\sum_{\specA\specB}Z_{\specA}Z_{\specB}
\bigg(\int\inftes{\vec{r}_1}\int\inftes{\vec{r}_2}\int\inftes{\vec{r}^{\,\prime}}\\
\times&\frac{\rho_{\specA}(\vec{r}_1)\rho_{\specB}(\vec{r}_2)
\delta\left(|\vec{r}^{\,\prime}-\vec{r}_1|-\DADBTwo\right)}
{\pi(\DADB)^2|\vec{r}_2-\vec{r}^{\prime}|}
\textrm{\,.}
\end{split}
\end{equation}
Moreover, similar to \cref{eqn:thetaderiv}, its functional derivative reads 
\begin{equation}\label{eqn:deltaderiv}
\begin{split}
\fdv{{\mathcal{F}}_{\textrm{ex}}^{\delta}\funcdepend}{\rho_{\specA}(\vec{r})}&=
\frac{Z_{\specA}}{2\beta}\sum_{\specB}\big(\phi_{\specB}\ast
\tilde{\delta}_{\specA\specB}\big)(\vec{r}) \\
&+\frac{Z_{\specA}}{2\beta}\sum_{\specB}\bjerrum\int\inftes{\vec{r}^{\,\prime}}
\frac{n_{\specA\specB}^{\delta}(\vec{r}^{\,\prime})}{|\vec{r}-\vec{r}^{\,\prime}|}
\textrm{\,.}
\end{split}
\end{equation}
Note that, as previously discussed for the last term of the functional itself, the derivative in \cref{eqn:deltaderiv} follows from \cref{eqn:thetaderiv} by skipping the last term. 
In bulk, \cref{eqn:deltaderiv} evaluates as shown in \cref{app:thetaderiv} such that the contributions to the modified chemical potentials from the $\functionaldelta$ functional are 
\begin{equation}
\label{eqn:mu-delta}
\begin{split}
\mu_{\specA}^{\ast\delta}&=
\left.\fdv{\functionaldelta\funcdepend}{\rho_{\specA}(\vec{r})}
\right|_{\rho_{\specA}(\vec{r})=\rho_{\specA}}\\
&=-\frac{\pi}{6}\frac{Z_{\specA}\bjerrum}{\beta}
\sum_{\specB}Z_{\specB}\rho_{\specB}(\DADB)^2
\textrm{\,.}
\end{split}
\end{equation}
These contributions vanish due to the charge neutrality in bulk if the hard-sphere diameters of the species are identical.

\section{Numerical Results}\label{sec:results}
\begin{figure}
\raggedright
\includegraphics[width=\linewidth]{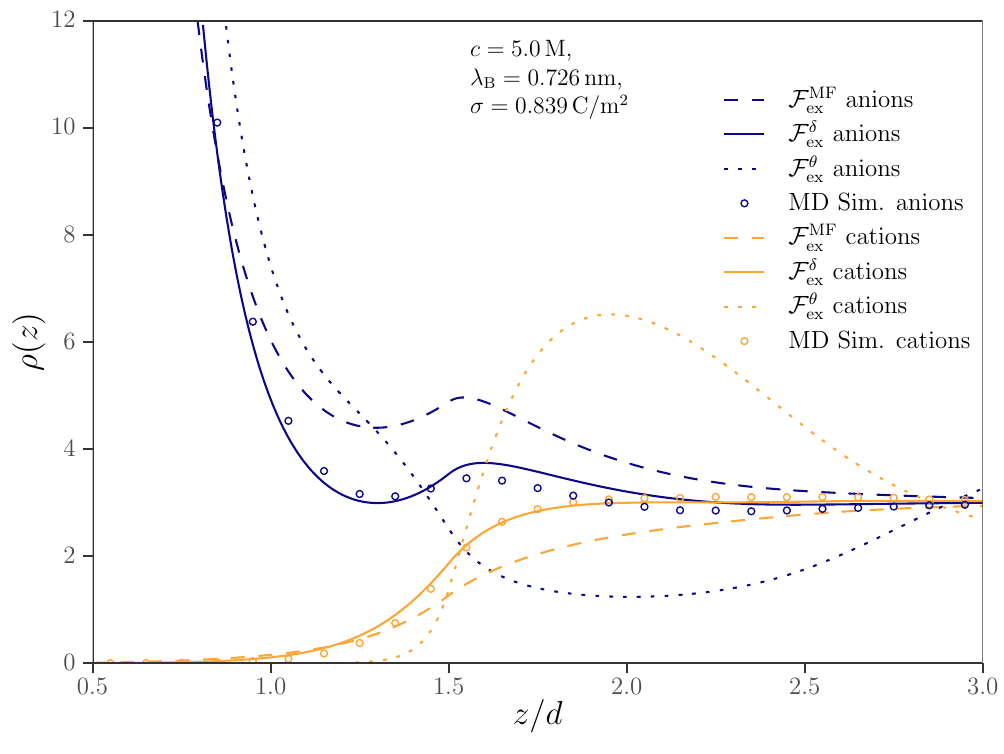}
\caption{
    Density profiles of negative (blue) and positive (yellow) ions with a hard-core diameter of \hbox{$d:=d_{\scriptscriptstyle+}=d_{\scriptscriptstyle-}$} predicted by $\functionalmf$ (dashed line), $\functionaldelta$ (solid line) and $\functionaltheta$ (dotted line) at a positively charged hard wall.
    For comparison, simulation results are shown as open circles.}
    \label{fig:thetaprofile}
\end{figure}
In this section we study (A) density profiles, (B) correlation functions, and (C) thermodynamic properties for the \PM{} that we compute with the aforementioned functionals.
Thus, we apply the mean-field functional $\functionalmf$ and the modified restricted phase space functional $\functionaldelta$ and compare results to molecular dynamics (\MD{}) simulation results obtained with the simulation package \texttt{ESPResSo~4.1.4}~\cite{weik2019espresso}.
Note that, apart from one example, we do not show results from the restricted phase space functional $\functionaltheta$, because its predictive capabilities are very limited:
For instance, we find contact values for a given electrostatic surface potential being overestimated in all calculations, or unrealistic layering effects occuring in the density profiles.
The latter is exemplary shown for the hard-wall geometry in \cref{fig:thetaprofile}.
Further, the functional showed numerical divergences in the minimizing Picard iteration scheme, hence hinting towards its low performance.

\begin{figure}
\raggedright
\includegraphics[width=.95\linewidth]{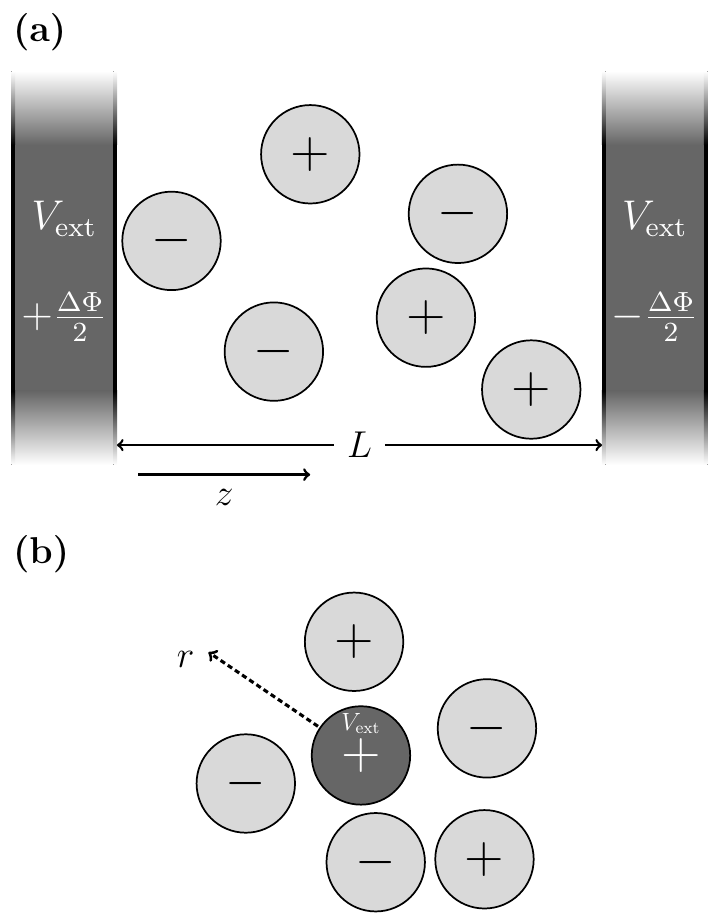}
\caption{
    Sketch of the two external potentials that we use in this work, also illustrating their induced geometries.
    (a) The setup with parallel charged hard walls, which is implemented in DFT via the external hard-wall potential $V_{\textrm{ext}}$.
    The walls are separated by a distance $L$ and have a potential difference of $\Delta\Phi$.
    (b) The test particle setup, where one particle is fixed and acts as an external potential $V_{\textrm{ext}}$ for the system.
}\label{fig:systems}
\end{figure}

For our study we implement two different geometries that are sketched in \cref{fig:systems}.
The first one is that of a system with two infinitely extended, hard walls in the $xy$-plane that are at distance $L$ to each other and have an electrostatic potential difference $\Delta\Phi$.
The respective external potential is expressed in \cref{eqn:extpotwalls}.
Due to the symmetries the density profiles only vary along the remaining Cartesian coordinate $z$ such that we obtain quasi one-dimensional density profiles $\rho_{\specA}(z)$.
The second one is the spherical symmetric geometry of the test particle setup, which we use to obtain total correlation functions in bulk.
Here, the respective external potential is expressed in \cref{eqn:extpotpercus}.
In this setup the one-body density profile around a fixed test particle is related to the two-body distribution \cite{percus1962approximation, frisch1964equilibrium}, as shown in \cref{eqn:percus}.
While the uniform bulk densities have no spatial dependence, the resulting pair distributions $g_{\specA\specB}^{(2)}(r)$ depend on the radial coordinate $r$. 
To close the system around the test particle, we add a hard, spherical outer wall with a distance of \hbox{$L=40\,\textrm{nm}$} to the origin.
This large separation $L$ ensures bulk behaviour in between the walls, even if the system resembles a spherical capacitor with the test particle in its center.

The calculation of the electrostatic potentials via the Poisson equation, the implementation of the external potential, and the computation of weighted densities
are explained in detail in \cref{app:numerical}.
The details of the \MD{} simulations are explained in \cref{app:simulation}.
%
\subsection{Density Profiles}\label{sec:results:profiles}
\begin{figure*}
\includegraphics[width = .99\textwidth]{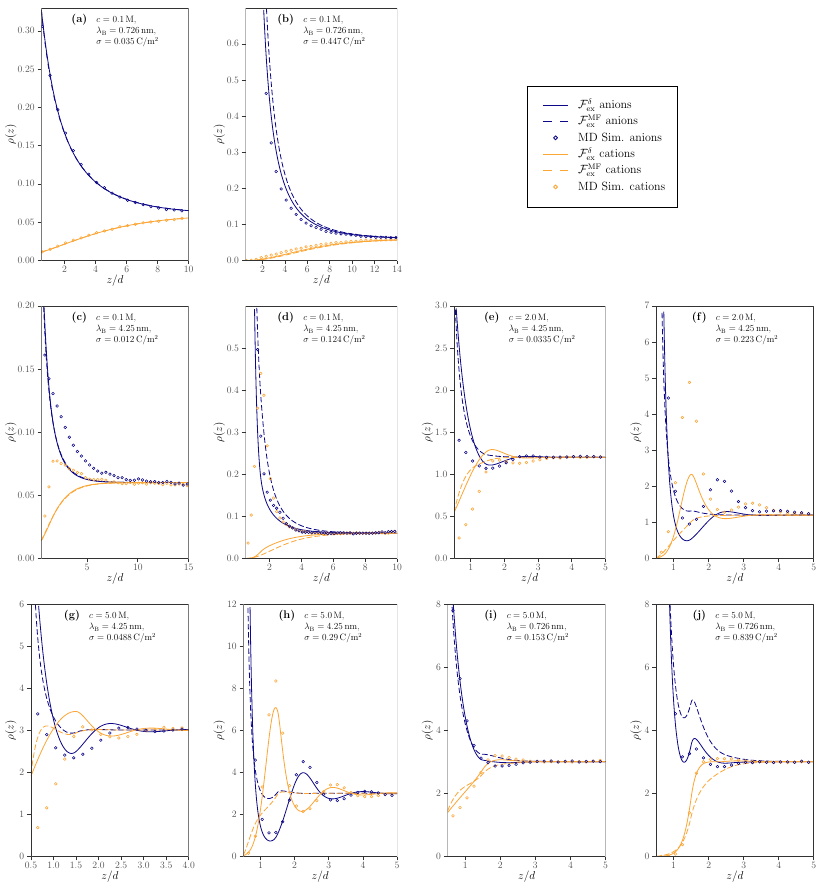}
\caption{
  Density profiles of charged hard spheres with a diameter of \hbox{$d:=d_{\scriptscriptstyle+}=d_{\scriptscriptstyle-}$} at a charged planar wall
  obtained from the $\functionaldelta$ functional (solid line) and the $\functionalmf$ functional (dashed line) from eqs.~\eqref{eqn:f-delta} and~\eqref{eqn:f-mf} for different concentrations and Bjerrum lengths.
  The parameters are given at the top of each panel and additionally are listed in~\cref{tab:parameterswalls}.
  Density profiles of negative ions are colored (dark) blue and density profiles of positive ions are colored (light) yellow.
  Data from \MD{} simulations is shown as open circles in the respective color for comparison.}
  \label{fig:densityprofile}
\end{figure*}

%
To test the accuracy of our $\functionaldelta$ functional in predicting density profiles, we calculate density profiles $\rho_{\specA}(z)$ on a fine numerical grid with a spacing of \hbox{$\Delta z=3\cdot10^{-5}\,\textrm{nm}$} (\hbox{$4\cdot10^5+1$ grid points}) in the aforementioned system with two planar hard walls and compared them to MD simulation results.
The two walls are \hbox{$L=12\,\textrm{nm}$} apart and have a fixed electrostatic potential difference of \hbox{$\Delta\Phi=0,0.1,0.5\,\textrm{V}$}, respectively. 
Moreover, we consider a range of concentrations \hbox{$c=0.05,\dots,5\,\textrm{M}$} and Bjerrum lengths \hbox{$\bjerrum=0.726,4.25\,\textrm{nm}$}.
The temperature is set to \hbox{$T=293.41\,\textrm{K}$}.
To show the shortcomings and predictive capabilities of our functional, we compare it to the standard mean-field functional $\functionalmf$ and \MD{} data with the same plate-surface charge density $\sigma$.
In \cref{fig:densityprofile} we show a representative set of density profiles.
All parameter combinations can be found in \cref{tab:parameterswalls}.

\begin{table*}[ht]
﻿\newcolumntype{Y}{D{.}{.}{4,6}}
\newcolumntype{V}{D{.}{.}{3,3}}
\begin{tabularx}{\textwidth}{lYYYYYY|YY|VY}
\toprule

\multirowcell{2}{figure\\label\,} &
\nodc{\multirowcell{2}{\\$d_{+}$}} &
\nodc{\multirowcell{2}{\\$d_{-}$}} &
\nodc{\multirowcell{2}{\\$c$}} &
\nodc{\multirowcell{2}{\\$\bjerrum$}} &
\nodc{\multirowcell{2}{\\$\sigma$}} &
\multicolumn{1}{c|}{\multirowcell{2}{\\$P$}} &
\multicolumn{2}{c|}{$\functionalmf$} &
\multicolumn{2}{c}{$\functionaldelta$}\\

&&&&&&&
\nodc{$\Delta\Phi$} &
\nodcrl{\CDT{}} &
\nodc{$\Delta\Phi$} &
\nodc{\CDT{}}\\

&
\nodc{[nm]} &
\nodc{[nm]} &
\nodc{[M]} &
\nodc{[nm]} &
\nodc{[$\textrm{C}/\textrm{m}^2$]} &
\multicolumn{1}{c|}{[$\boltzmann\!T\!/\textrm{nm}^3$]} &
\nodc{[V]} &
\multicolumn{1}{c|}{[\%]} &
\nodc{[V]} &
\nodc{[\%]}\\
\midrule

\ref*{fig:densityprofile}(a) & 0.3 & 0.3 & 0.1 & 0.726 & 0.035  & 0.121 & 0.1009 & \nodcrl{$<0.002$}     & 0.1 &\nodc{$<0.002$}\\ 
\ref*{fig:densityprofile}(b) & 0.3 & 0.3 & 0.1 & 0.726 & 0.447  & 0.121 & 0.5469 & \nodcrl{$<0.2$}       & 0.5 &\nodc{$<0.3$}\\ 
\ref*{fig:densityprofile}(c) & 0.3 & 0.3 & 0.1 & 4.25  & 0.012  & 0.121 & 0.1035 & \nodcrl{$<0.02$}      & 0.1 &\nodc{$<0.02$}\\ 
\ref*{fig:densityprofile}(d) & 0.3 & 0.3 & 0.1 & 4.25  & 0.124  & 0.121 & 0.605  & \nodcrl{$<0.1^{\,*}$} & 0.5 &\nodc{$<0.2^{\,*}$}\\ 
\ref*{fig:densityprofile}(e) & 0.3 & 0.3 & 2.0 & 4.25  & 0.0335 & 2.767 & 0.1305 & \nodcrl{$<0.02$}      & 0.1 &\nodc{$<0.02$}\\ 
\ref*{fig:densityprofile}(f) & 0.3 & 0.3 & 2.0 & 4.25  & 0.223  & 2.767 & 0.7649 & \nodcrl{$<1.1$}       & 0.5 &\nodc{$<2$}\\ 
\ref*{fig:densityprofile}(g) & 0.3 & 0.3 & 5.0 & 4.25  & 0.0488 & 8.586 & 0.1621 & \nodcrl{$<0.02$}      & 0.1 &\nodc{$<0.03$}\\ 
\ref*{fig:densityprofile}(h) & 0.3 & 0.3 & 5.0 & 4.25  & 0.29   & 8.586 & 0.9079 & \nodcrl{$<3.3$}       & 0.5 &\nodc{$<1.4$}\\ 
\ref*{fig:densityprofile}(i) & 0.3 & 0.3 & 5.0 & 0.726 & 0.153  & 8.586 & 0.1179 & \nodcrl{$<0.03$}      & 0.1 &\nodc{$<0.03$}\\ 
\ref*{fig:densityprofile}(j) & 0.3 & 0.3 & 5.0 & 0.726 & 0.839  & 8.586 & 0.6077 & \nodcrl{$<2.4$}       & 0.5 &\nodc{$<1.5$}\\ 
\bottomrule
\end{tabularx}\\
\vspace{1em}
\newcolumntype{W}{D{.}{.}{8,8}}
\newcolumntype{U}{D{.}{.}{4,6}}
\begin{tabularx}{\textwidth}{UUUUUU|WU|WU}
\toprule

\nodc{\multirowcell{2}{\\$d_{+}$}} &
\nodc{\multirowcell{2}{\\$d_{-}$}} &
\nodc{\multirowcell{2}{\\$c$}} &
\nodc{\multirowcell{2}{\\$\bjerrum$}} &
\nodc{\multirowcell{2}{\\$\Delta\Phi$}} &
\multicolumn{1}{c|}{\multirowcell{2}{\\$P$}} &
\multicolumn{2}{c|}{$\functionalmf$} &
\multicolumn{2}{c}{$\functionaldelta$}\\

&&&&&&
\nodc{$\sigma$} &
\nodcrl{\CDT{}} &
\nodc{$\sigma$} &
\nodc{\CDT{}}\\

\nodc{[nm]} &
\nodc{[nm]} &
\nodc{[M]} &
\nodc{[nm]} &
\nodc{[V]} &
\multicolumn{1}{c|}{[$\boltzmann\!T\!/\textrm{nm}^3$]} &
\nodc{[$\textrm{C}/\textrm{m}^2$]} &
\multicolumn{1}{c|}{[\%]} &
\nodc{[$\textrm{C}/\textrm{m}^2$]} &
\nodc{[\%]}\\
\midrule
 0.25  & 0.3  & 5.0  & 4.25   & 0.0  & 7.824  & \nodc{~-0.0037,\,-0.0037~} & \nodcrl{$0.01$} & \nodc{~0.0092,\,-0.0092~}&\nodc{$0.6$}\\
 0.25  & 0.3  & 5.0  & 4.25   & 0.1  & 7.824  & \nodc{~0.0278,\,-0.0361~}  & \nodcrl{$0.03$} & \nodc{~0.0389,\,-0.0590~}&\nodc{$0.6$}\\ 
 0.25  & 0.3  & 5.0  & 4.25   & 0.5  & 7.824  & \nodc{~0.154,\,-0.178~}    & \nodcrl{$0.7$}  & \nodc{~0.274,\,-0.331~}  &\nodc{$0.6$}\\ 
\bottomrule
\end{tabularx}\caption{
Overview over the parameters we use in our DFT calculations of the \PM{} at charged hard walls. The panels where the respective density profiles are shown are referred to in the first column.
The temperature $T=293.41\,\textrm{K}$ is the same for all systems.
$d_{\scriptscriptstyle+}$ and $d_{\scriptscriptstyle-}$ are the hard diameters of the respective ions, $c$ is the concentration, $\bjerrum$ is the Bjerrum length, $\Delta\Phi$ is the potential difference between the two electrode walls, $\pm\sigma$ is the surface charge density on the respective wall, $P$ is the pressure calculated from the grand potential, and ``\CDT{}'' refers to the relative deviation between the left-hand side (\lhs{}) and the right-hand side (\rhs{}) of the contact density theorem, \cref{eqn:cdt}, via \hbox{$|((\textrm{\lhs})-(\textrm{\rhs}))/(\textrm{\lhs})|$}.
The '$<$' indicates that the corresponding CDT difference would decrease significantly for higher grid resolutions.
The '*' means that the system length $L$ is extended to $20\,\textrm{nm}$, because the density profiles decay very weakly.}
\label{tab:parameterswalls}
\end{table*}

Before we discuss the shown density profiles, we have a closer look on uncharged walls.
In the \PM{}, we expect a so called \emph{depletion layer}, which is a negative deviation from the bulk densities close to the walls, which affects all particle species equally.
It stems from electrostatic screening that an ion experiences from the other ions in the bulk, but not from the walls itself.
Hence, ions that are close to the wall feel a net repulsion from the wall and into the direction of bulk which leads to a drop in the density profiles close to the wall.
However, it follows from \cref{eqn:f-mf,eqn:f-delta} that in the absence of a potential difference $\Delta\Phi$ our functionals vanish, which is due to the external potential being the only driver of a difference in the density profiles of the differently charged ions.
If the potential difference is set to zero, the modulus of all terms in the sums in \cref{eqn:f-mf,eqn:f-delta} become equal, such that the alternating sign due to the valencies makes the functionals vanish.
The remaining hard-sphere functional does not capture the expected depletion layer of the \PM{} at the hard walls.
In other works functionals were developed that capture this effect qualitatively~\cite{roth2016shells}.

The depletion effect is much weaker than electrostatic effects from the charged wall, which leads to its suppression as soon as we apply a small potential difference \hbox{$\Delta\Phi=0.1\,\textrm{V}$}.
For low concentrations (\hbox{$c=0.1\,\textrm{M}$}) and weak electrostatic interactions (\hbox{$\bjerrum = 0.726\,\textrm{nm}$}) both the $\functionalmf$ and $\functionaldelta$ functionals show the same behavior and agree with the simulation data, as shown in \cref{fig:densityprofile}(a).
Increasing the potential difference to \hbox{$\Delta\Phi=0.5\,\textrm{V}$}, which is a typical value applied in supercapacitors, we can see from 
\cref{fig:densityprofile}(b) that the density profile of $\functionaldelta$ fits the simulation results more closely than the ones from $\functionalmf$, but the differences are rather small.
To show qualitative differences between the two functionals, we increase the electrostatic interactions such that \hbox{$\bjerrum=4.25\,\textrm{nm}$}.
As one can see in \cref{fig:densityprofile}(c)~and~(d), now both functionals fail to match the simulation data in this situation, because a peak of co-ions appears close to the charged walls at $z\approx2\,d$ in the density profiles of the simulation data, which is not predicted by the functionals (for convenience we use \hbox{$d:=d_{\scriptscriptstyle+}=d_{\scriptscriptstyle-}$}). 
This discrepancy can be explained by regarding the phase diagram of the \PM{} as depicted in \cite{fantoni2013monte}.
The critical point for this phase diagram is roughly situated at $\rho^{\ast}\approx2.5\cdot10^{-2}$ and $T^{\ast}=d/\bjerrum\approx0.053$.
Our systems with \hbox{$\bjerrum=4.25\,\textrm{nm}$}, which corresponds to a reduced temperature of $T^{\ast}\approx0.07$, are much closer to the triple point than the ones with \hbox{$\bjerrum=0.726\,\textrm{nm}$} ($T^{\ast}\approx0.4$).
Close to the critical point correlation lengths increase, which for instance leads to the formation of \hbox{2--1} open ion clusters and \hbox{2--2} tetrameters~\cite{orkoulas1994free}.
These clusters allow the accumulation of co-ions at a distance of \hbox{$z\leq5\,d$} from the wall, which results in a peak in the respective density profile.
To our knowledge, there have not been any attempts to find a functional that captures these clustering effects yet.

Now, we study our system at higher concentrations, a situation where the mean-field functional and other approaches typically become less predictive or break down.
This is not surprising, because the mean-field functional is deduced from a low-density approximation.
Interestingly, this is not the case for our $\functionaldelta$ functional.
As one can see from \cref{fig:densityprofile}(e) and (f), at an intermediate concentration of \hbox{$c=2\,\textrm{M}$} $\functionaldelta$ matches the simulation at a qualitative level by showing similar layering effects, but with the wrong peak height.
In contrast, the regular mean-field functional, does not show this layering behavior.
Even at a concentration of \hbox{$c=5\,\textrm{M}$} its density profile does not show any layering, whereas the density profiles of $\functionaldelta$ match the simulation data closely and predict the pronounced layering effects shown in \cref{fig:densityprofile}(g)--(j) well.

\begin{figure}
\includegraphics[width = .99\linewidth]{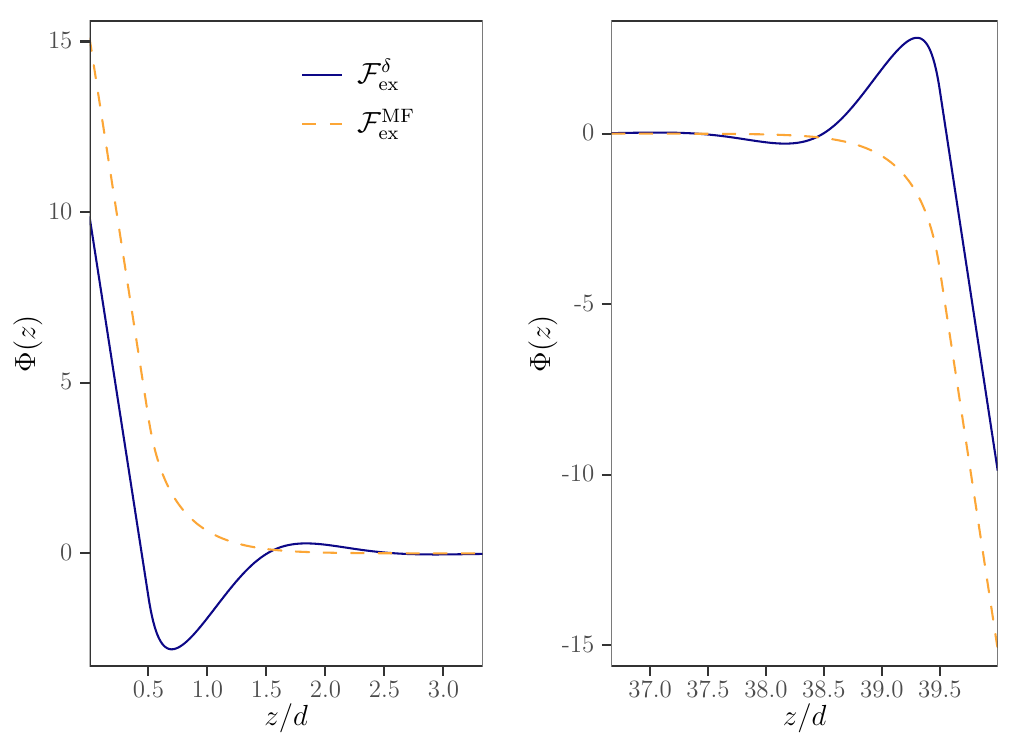}
\caption{
  Dimensionless total electrostatic potential $\Phi(z)$ in a system of charged hard spheres with a diameter of \hbox{$d:=d_{\scriptscriptstyle+}=d_{\scriptscriptstyle-}$} at a concentration \hbox{$c=2\,\textrm{M}$} between two planar charged hard walls with \hbox{$\sigma=0.223\,\textrm{C}/\textrm{m}^2$}. 
  The Bjerrum length is \hbox{$\bjerrum=4.25\,\textrm{nm}$}.
  Respective density profiles are shown in \cref{fig:densityprofile}(f).
  We show results for the functionals $\functionaldelta$ (blue, solid line) and $\functionalmf$ (yellow, dashed line) from eqs.~\eqref{eqn:f-delta} and~\eqref{eqn:f-mf}.}
  \label{fig:chargeinversion}
\end{figure}

These pronounced layering effects also have a considerable influence on the total electrostatic potential \hbox{$\Phi(z)$} of the system.
As \cref{fig:chargeinversion} shows for the density profiles shown in \cref{fig:densityprofile}(f), the regular mean-field functional generates a monotonic potential, whereas the modified version does show two distinct extrema close to the hard wall.
The change of sign in the slope of the potential reveals that the electric field is not only weakened by the layering effects but actually is reversed.
Moreover, from the density profiles one can see that the net charge in some regions close to the positive electrode is negative and vice versa.
This phenomenon is called charge inversion~\cite{van1980grand} and is relevant for example in the transition of differential capacitance curves from bell-shaped to camel-shaped~\cite{kornyshev2007double}.
Thus, predicting this charge-inversion effect is an important property that the regular mean-field 
functional lacks.

Of course, other functionals exist that are also more involved than the mean-field functional and also have more predictive power.
We now focus on two functionals derived from the mean spherical approximation (\MSA{}), which follows from a closure relation for the \OZ{} equation in liquid state theory.
The \MSA{} closure respects the volume exclusion of hard spheres in the total pair-correlation function by prescribing its inner part to vanish.
Furthermore, the direct correlation function is set to be equal to the tail of an additional soft pair potential (with the additional prefactor $-\beta$) for radial positions exceeding the contact distance. 
The solution of the OZ equation with the \MSA{} closure for an electrostatic pair potential's tail was first derived by~\citeauthor{waisman1972mean}~\cite{waisman1972mean}. From this solution \citeauthor{roth2016shells} constructed the MSAc functional~\cite{roth2016shells}.
The MSAc functional, however, does not predict the correct bulk free energy.
The authors resolved this problem in the MSAu functional, where the bulk free-energy solution from the \MSA{} is added to the MSAc functional and the bulk densities in this additional term are simply replaced by weighted density profiles. 
As a consequence, the MSAu functional is capable of predicting depletion at uncharged walls in contrast to all other approaches. 
We now compare our $\functionaldelta$ and $\functionalmf$ functionals to results from ref.~\cite{cats2021primitive} that are obtained by these MSAc and MSAu functionals in a study that focuses on the decay of correlation functions.
For this purpose, we introduce the dimensionless charge density $\rho_{\textrm{Z}}$ and number density $\rho_{\textrm{N}}$, defined as
\begin{equation}\label{eqn:ccdddensities}
\begin{split}
\rho_{\textrm{Z}}(z)&=
\frac{\rho_{+}(z)-\rho_{-}(z)}{\rho_{\textrm{b}}}\\
\rho_{\textrm{N}}(z)&=
\frac{\rho_{+}(z)+\rho_{-}(z)}{\rho_{\textrm{b}}}-2
\textrm{\,.}
\end{split}
\end{equation}
Each of these densities reveals a distinct decay length~\cite{coupette2018screening}, which recently attracted attention due to the results of surface force measurements~\cite{gebbie2013ionic,smith2016electrostatic} showing the so called underscreening effect.
This effect represents a strong increase of the decay length with increasing concentration and could not yet be explained theoretically \cite{coles2020correlation,kjellander2020multiple,zeman2020screening,cats2021primitive}.

\begin{figure}
\includegraphics[width = .99\linewidth]{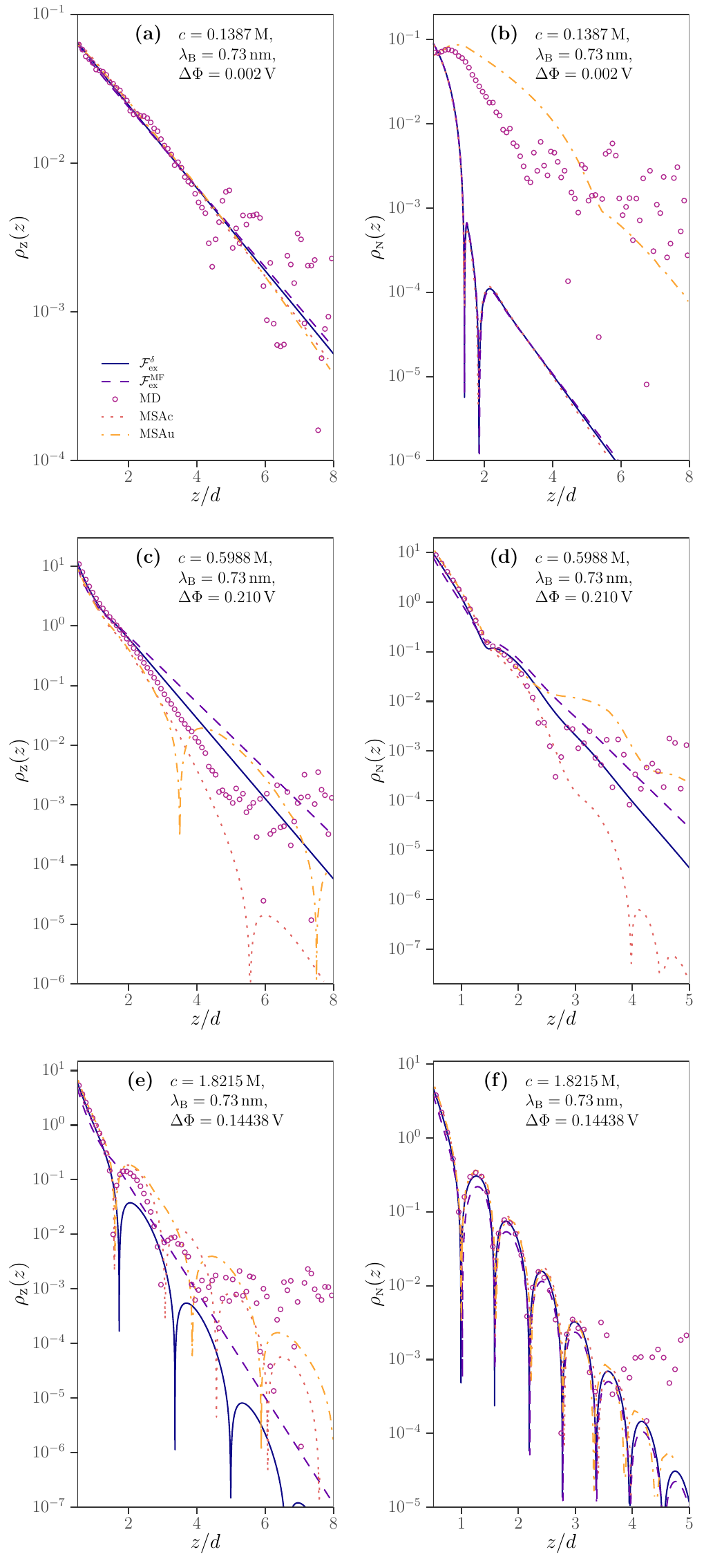}
\caption{
Dimensionless charge and number density profiles $\rho_{\textrm{\tiny Z}}$ (left column) and $\rho_{\textrm{\tiny N}}$ (right column) of counterions in the vicinity of a charged hard wall in the \PM{}.
We show results from the $\functionaldelta$ (blue, solid line) and $\functionalmf$ (violet, dashed line) functionals as well as from the MSAc (red, dotted line) and MSAc (yellow, dot-dashed line) functionals and \MD{} simulation results (purple open circles). Data for MSAc, MSAu, and \MD{} are from ref.~\cite{cats2021primitive}. Data for $\functionalmf$ correspond to that of ${\mathcal{F}}_{\textrm{ex}}^{\textrm{MFC}}$ in ref.~\cite{cats2021primitive} (both mean-field functionals are identical).
The hard-sphere diameter and Bjerrum length are set to \hbox{$d:=d_{\scriptscriptstyle+}=d_{\scriptscriptstyle-}=0.5\,\textrm{nm}$} and \hbox{$\bjerrum=0.73\,\textrm{nm}$} in all panels. The respective concentration and applied electrostatic potential are noted on the top of each panel.}
\label{fig:comparisonroth}
\end{figure}

\begin{table*}
\newcolumntype{Z}{D{.}{.}{5,6}}
\begin{tabularx}{\textwidth}{XZZZZZZZZ}
\toprule

\multirowcell{2}{figure~~~~~~\\\!label~~~~~~} &
\nodc{\multirowcell{2}{\\$d_{\scriptscriptstyle+}$}} &
\nodc{\multirowcell{2}{\\$d_{\scriptscriptstyle-}$}} &
\nodc{\multirowcell{2}{\\$c$}} &
\nodc{\multirowcell{2}{\\$\bjerrum$}} &
\nodc{\multirowcell{2}{\\$P$}} &
\nodc{\multirowcell{2}{Simulation\\Virial}} &
\nodc{\multirowcell{2}{$\functionalmf$\\Virial}} &
\nodc{\multirowcell{2}{$\functionaldelta$\\Virial}}\\

&&&&&&&&\\

&
\nodc{[nm]} &
\nodc{[nm]} &
\nodc{[M]} &
\nodc{[nm]} &
\nodc{[$\boltzmann\!T\!/\textrm{nm}^3$]} &
\nodc{[$\boltzmann\!T\!/\textrm{nm}^3$]} &
\nodc{[$\boltzmann\!T\!/\textrm{nm}^3$]} &
\nodc{[$\boltzmann\!T\!/\textrm{nm}^3$]}\\
\midrule

\ref*{fig:totalcorrelations}(a)   & 0.3  & 0.3  & 0.1   & 3.0  & 0.121 & 0.058(3) & 0.053 & 0.053 \\
\ref*{fig:totalcorrelations}(b)   & 0.3  & 0.3  & 3.0   & 3.0  & 4.189 & 1.46(13) & 1.569 & 1.536 \\
\ref*{fig:totalcorrelations}(c)   & 0.3  & 0.3  & 10.0  & 3.0  & 18.76 & 12.1(7)  & 13.64 & 13.87 \\
\ref*{fig:totalcorrelations}(d)   & 0.3  & 0.3  & 10.0  & 0.3  & 18.76 & 23.95(4) & 22.75 & 22.71 \\
\bottomrule
\end{tabularx}
\caption{
Overview over the parameters we use in our DFT calculations for the \PM{} in the test particle setup, as discussed in \cref{sec:results:correlation}.
The panels where the respective total-correlation functions are shown are referred to in the first column.
The temperature is the same for all systems with  $T=293.41\,\textrm{K}$.
$d_{\scriptscriptstyle+}$ and $d_{\scriptscriptstyle-}$ are the hard-sphere diameters of the ions, $c$ is the concentration, $\bjerrum$ is the Bjerrum length, $P$ is the pressure calculated from the grand potential, and 'Virial' refers to the right-hand side (\rhs{}) of the virial pressure equation~\eqref{eqn:virialions}.}
\label{tab:parametersspherical}
\end{table*}

In \cref{fig:comparisonroth} we show the density profiles of the aforementioned functionals on a semi-logarithmic scale.
This scaling allows to determine the decay length immediately from the curves' slopes.
Panel (a) shows that all charge-density profiles coincide with the simulation data at low concentrations and weak electrostatic interactions and potentials.
Panel (b) shows the respective number-density profiles.
Since the potential difference is almost zero, the excess number densities are governed by the fact whether the functional captures the repulsion from an (almost) uncharged wall.
The MSAu functional is the only one that can predict this effect even if it does so rather inaccurately.
The drawback of the MSAu functional however seems to be the overall predictive capability, because it shows oscillations in panel (c) and an inaccurate decay in panel (d) that do not appear in the simulation data.
Even though the MSAc functional predicts the simulation data quite well in panels (c) and (d), its density profiles show strange oscillatory behavior at \hbox{$z>4\,d$}.
The accuracy of the prediction of the modified mean-field functional $\functionaldelta$ is comparable to that of the MSAc functional, but compared to MSAc and MSAu it does not show oscillatory behavior, which is in agreement with the simulations.
However, in panel (e) one can see that it slightly overestimates the decay strength.
Note, that the mean-field functional always underestimates the strength of the decay and is not able to predict the oscillations in the charge density profiles at all.
In panel (f) the excess density profiles of all functionals agree with the simulation data, a fact that does not surprise, because all functionals use a very accurate \FMT{} functional for the hard-sphere interactions that dominate the number densities at high concentrations.

In summary, the $\functionaldelta$ functional as well as the $\functionalmf$ functional predict density profiles between parallel hard walls well at low densities and when the electrostatic interaction strength resembles a salt solution in water (\hbox{$\bjerrum=0.726\,\textrm{nm}$}), but only as long as an external electrostatic field is applied.
If the external field is missing one expects a depletion layer close to the walls, which both functionals do not predict because their respective electrostatic contributions vanish completely.
Furthermore, they cannot predict the clustering effects that arise at low concentrations \hbox{$c=0.1\,\textrm{M}$} and high Bjerrum lengths \hbox{$\bjerrum=4.25\,\textrm{nm}$}.
However, the $\functionaldelta$ functional performs well at high concentrations (\hbox{$>2\,\textrm{M}$}) and when high potential differences (\hbox{$>0.1\,\textrm{V}$}) are applied to the charged walls.
It also shows great improvements to decay behavior predictions with respect to the mean-field functional.
Compared to other sophisticated functionals like the aforementioned MSAc and MSAu it predicts decay behavior equally well.
%
\subsection{Correlation Functions}\label{sec:results:correlation}
%
\begin{figure*}
\includegraphics[width = .99\textwidth]{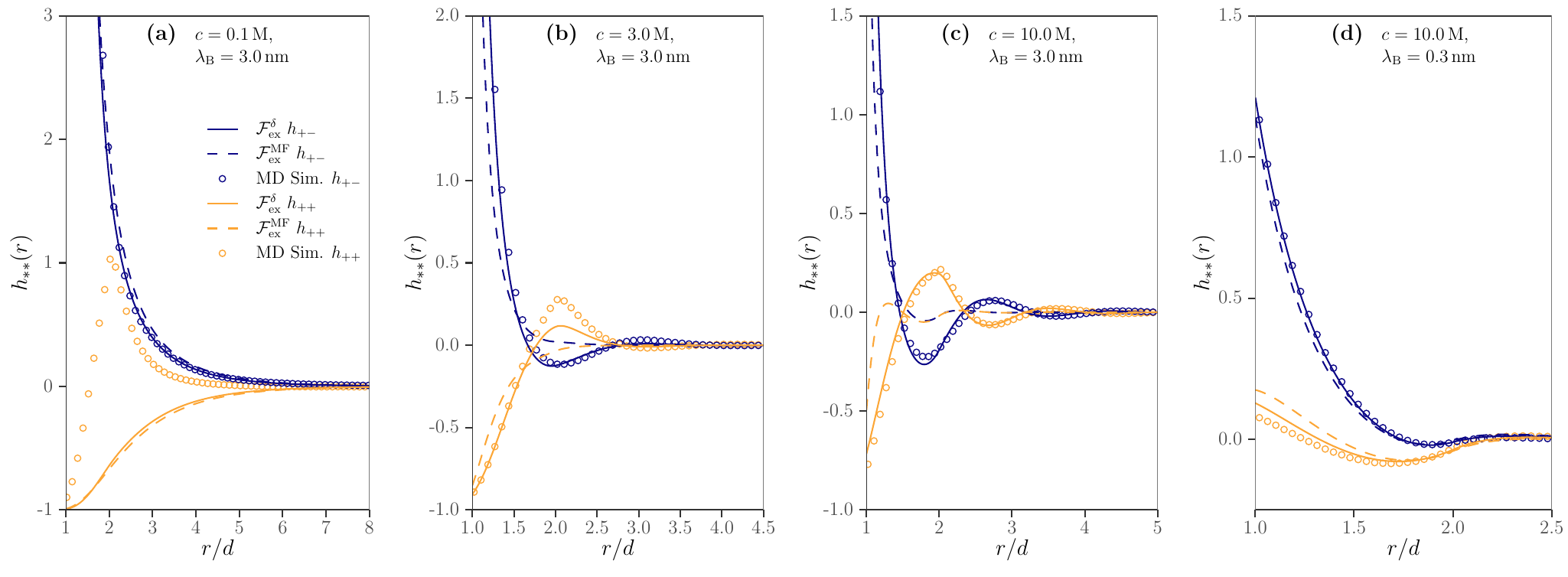}
\caption{
Total correlation functions of charged hard spheres with a diameter of \hbox{$d:=d_{\scriptscriptstyle+}=d_{\scriptscriptstyle-}$} obtained via the Percus trick from the functionals $\functionaldelta$ (solid line) and $\functionalmf$ (dashed line) for different concentrations and Bjerrum lengths.
The parameters are given at the top of each panel and are listed in \cref{tab:parametersspherical}.
The \hbox{\raisebox{.15em}{\tiny$++$} and \raisebox{.15em}{\tiny$--$} correlations} are identical and are colored orange.
The same holds for the \hbox{\raisebox{.15em}{\tiny$+-$} and \raisebox{.15em}{\tiny$-+$} correlations} which are colored blue.
Data from \MD{} simulations is shown as open diamonds in the respective color for comparison.}
\label{fig:totalcorrelations}
\end{figure*}
Now, we have a closer look on the bulk total correlation functions generated by the functionals $\functionalmf$ and $\functionaldelta$. We derive the correlation functions via \cref{eqn:percus}, thus, using the density profiles of test particle systems.
These systems have a radial size of \hbox{$L=40\,\textrm{nm}$} and a fixed temperature of \hbox{$T=293.41\,\textrm{K}$}.
We vary the particle concentrations \hbox{$c=0.1,3,10\,\textrm{M}$} and the Bjerrum length \hbox{$\bjerrum=0.3,3\,\textrm{nm}$}.
The density profiles are computed on a fine numerical grid with a spacing of \hbox{$\Delta r=4\cdot10^{-4}\,\textrm{nm}$} (\hbox{$10^5+1$ grid points}). 
All parameter combinations can be found in \cref{tab:parametersspherical}.
The obtained total correlation functions are shown in \cref{fig:totalcorrelations}.

In \cref{sec:results:profiles} we investigated density profiles between two parallel walls with peaks close to the walls due to ion clustering at low concentrations and high Bjerrum lengths.
It comes as no surprise that these clustering peaks also occur in the total correlation functions, because the latter can be related to the density profiles of the planar geometry. This correspondence is, for example, relevant in the context of decay lengths~\cite{cats2021primitive}.
A more intuitive explanation of the mentioned peak is possible on the basis of pair distributions: The repulsion between like-charge ions in a cluster as, for instance, the 2--2 neutral tetrameter, is essentially canceled by the other ions in the cluster.
The latter leads to a peak at \hbox{$r\approx2\,d$} in the like-charge total correlation function for low concentrations (\hbox{$c=0.1\,\textrm{M}$}) and high Bjerrum length (\hbox{$\bjerrum=3\,\textrm{nm}$}), as we can see in \cref{fig:totalcorrelations}(a).
Again, both functionals $\functionalmf$ and $\functionaldelta$ do not produce density profiles that show this effect.
When we increase the concentration, the modified functional $\functionaldelta$ predicts the occurring layering effects well as shown in \cref{fig:totalcorrelations}(b),(c), while the regular mean-field functional $\functionalmf$ produces inaccurate total correlation functions.
Here, the different curves for \hbox{\raisebox{.15em}{\tiny$++$} and \raisebox{.15em}{\tiny$+-$}} correlations seem to ``stick together'', an effect arising from the mean-field description of the Coulomb interaction that overestimates attraction of oppositely charged particle species at one and the same position when the hard-core repulsion is ignored.
This effect obviously is cured by the functional $\functionaldelta$.
At high concentrations (\hbox{$c=10\,\textrm{M}$}) but low electrostatic interaction strengths (\hbox{$\bjerrum=0.3\,\textrm{nm}$}) as shown in \cref{fig:totalcorrelations}(d) both functionals predict similar total correlation functions, but the modified mean-field functional still agrees better with the simulation data, in particular if we compare contact values.

As mentioned earlier, the decay behavior can be investigated not only in the total correlation functions obtained via the Percus trick in \cref{eqn:percus} but also in those obtained via the \OZ{} equation from the direct correlation functions of \cref{eqn:directcorrelation}. 
Many phenomena like wetting, clustering, or percolation are affected by the long-range decay of correlations.
Its theoretical description has been questioned in recent measurements by the phenomenon of underscreening~\cite{perez2017underscreening}, as discussed along \cref{eqn:ccdddensities}.
As for the density profiles in \cref{eqn:ccdddensities}, one defines charge (CC) and number (DD) correlations by~\cite{hansen2013theory_splitting}
\begin{equation}\label{eqn:correlationsplitting}
\begin{split}
h_{\textrm{CC}}(r)&=
\big(h_{\scriptscriptstyle++}(r)-h_{\scriptscriptstyle+-}(r)\big)\\
h_{\textrm{DD}}(r)&=
\frac{1}{2}\big(h_{\scriptscriptstyle++}(r)+h_{\scriptscriptstyle+-}(r)\big)
\textrm{\,,}
\end{split}
\end{equation}
because this choice reflects the eigenvectors $(-1,1)$ and $(1,1)$ from the matrix representation of the respective \OZ{} equations (see \cref{eqn:oz-matrix}) for a completely symmetric system like the \PM{} with equally sized ions.
Note that the prefactors in \cref{eqn:correlationsplitting} can be chosen freely and we follow the definition from \cite{hansen2013theory_splitting}.
The total correlation functions are determined from the Fourier transformed \OZ{} equations for multiple species that can be combined in a matrix notation by
\begin{equation}\label{eqn:oz-matrix}
\begin{split}
\hat{\textrm{H}}^{(2)}(k)
&\equiv\begin{pmatrix}\hat{h}_{++}^{(2)}(k) & \hat{h}_{+-}^{(2)}(k)\\ \hat{h}_{-+}^{(2)}(k) & \hat{h}_{--}^{(2)}(k)\end{pmatrix} \\
&=
\left(\mathbbm{1}-\hat{\textrm{C}}^{(2)}(k)\bar{\rho}_{\textrm{b}}\right)^{-1}
\hat{\textrm{C}}^{(2)}(k)
\textrm{\,.}
\end{split}
\end{equation}
The direct-correlation matrix $\hat{\textrm{C}}^{(2)}$ is defined equivalently.
The density matrix $\bar{\rho}_{\textrm{b}}$ contains the bulk densities of positive and negative particles on its diagonal and is zero elsewhere.
The asymptotic behavior of a total correlation function $h_{\specA\specB}^{(2)}$ is determined from the poles of the respective component of the rhs of \cref{eqn:oz-matrix}, more precisely by the pole with smallest imaginary part (see ref.~\cite{coupette2018screening} and references therein). 
This pole leads to an exponential decay of \hbox{$r\cdot h_{CC}(r)$} and \hbox{$r\cdot h_{DD}(r)$}, respectively.
Depending on the real part of the respective complex-valued poles, these decays can additionally show oscillations.

\begin{figure}
\includegraphics[width = .99\linewidth]{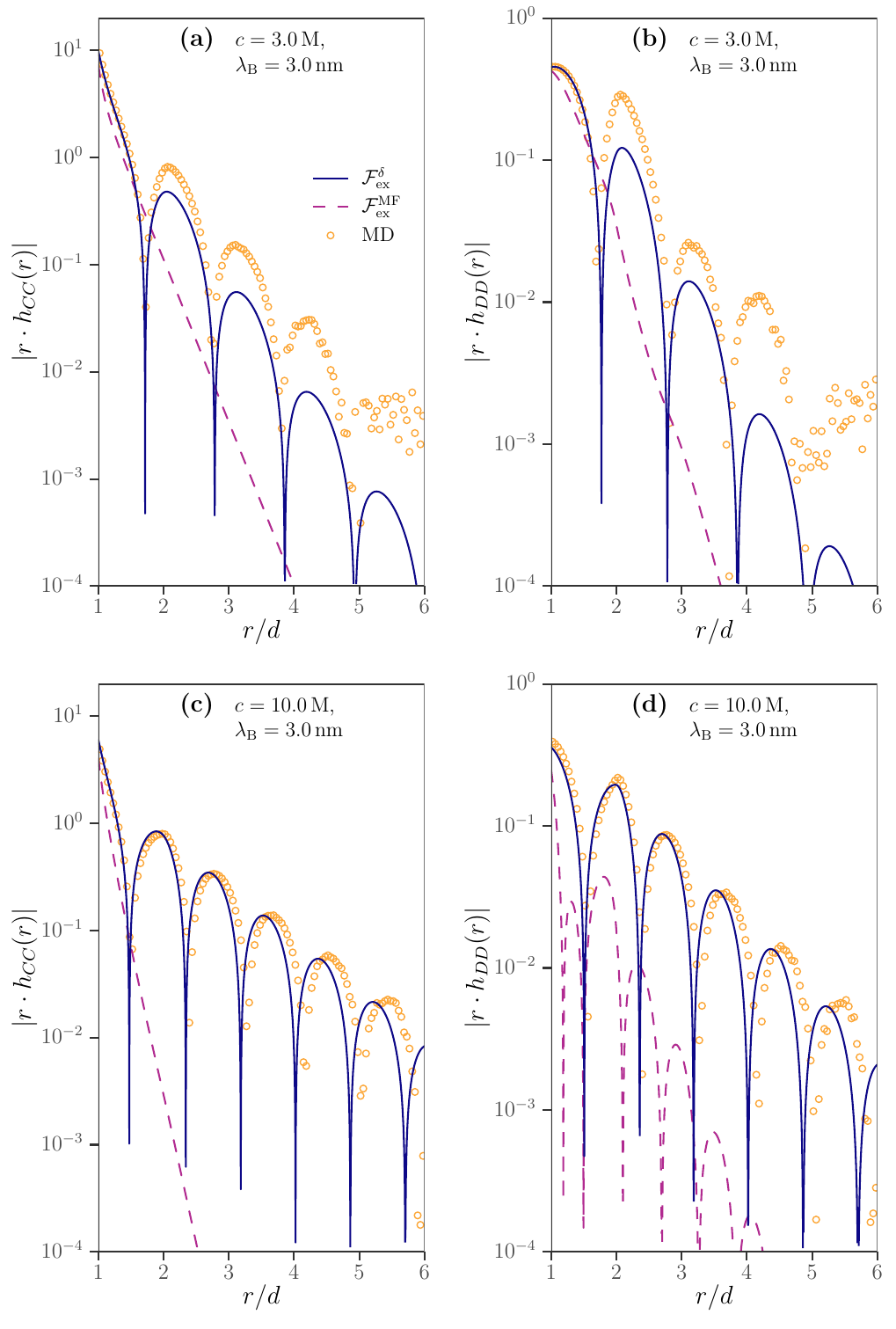}
\caption{
  The charge correlations (a), (c) and number correlations (b), (d) for charged hard spheres with a diameter of \hbox{$d:=d_{\scriptscriptstyle+}=d_{\scriptscriptstyle-}$}.
  The respective parameters are noted on the top of each panel and are listed in \cref{tab:parametersspherical} (b) and (c).
  The standard mean-field functional (dashed, red line) and the $\functionaldelta$ functional $\functionaldelta$ (solid, blue line) are compared with \MD{} simulation data (yellow, open circles).}
  \label{fig:ccddcorrelations}
\end{figure}

In \cref{fig:ccddcorrelations} the charge and number correlation functions are shown for two different concentrations \hbox{$c=3,10\,\textrm{M}$}.
For \hbox{$c=3\,\textrm{M}$}, panels (a) and (b) of \cref{fig:ccddcorrelations} show that the charge and the number correlations from $\functionalmf$ do not exhibit any oscillations and the decay length is greatly underestimated, whereas the correlations of $\functionaldelta$ show the correct oscillation frequency and just a slightly lower decay length as the simulation data.
For the higher concentration of \hbox{$c=10\,\textrm{M}$}, panels (c) and (d) of \cref{fig:ccddcorrelations} show that the functional $\functionalmf$ still predicts no oscillations in the charge correlation function and an inaccurate oscillation in the number correlation function.
Moreover, it still underestimates the decay length in both cases.
The $\functionaldelta$ correlation functions agree very well with the simulation data.

An interesting method to study the quality of an energy functional and its deduced entities like the density profiles and correlation functions is to test thermodynamic consistency.
In the last paragraphs we have studied correlation functions that we obtained, on one hand, via the Percus trick and, on the other hand, via functional derivatives and the \OZ{} equation.
For the exact functional these routes, the test-particle route and the thermodynamic or compressibility route, must be consistent and must yield the same pair-correlation functions.
Thus, we can test a functional for consistency by comparing correlations calculated via both routes.

\begin{figure}
\includegraphics[width = .99\linewidth]{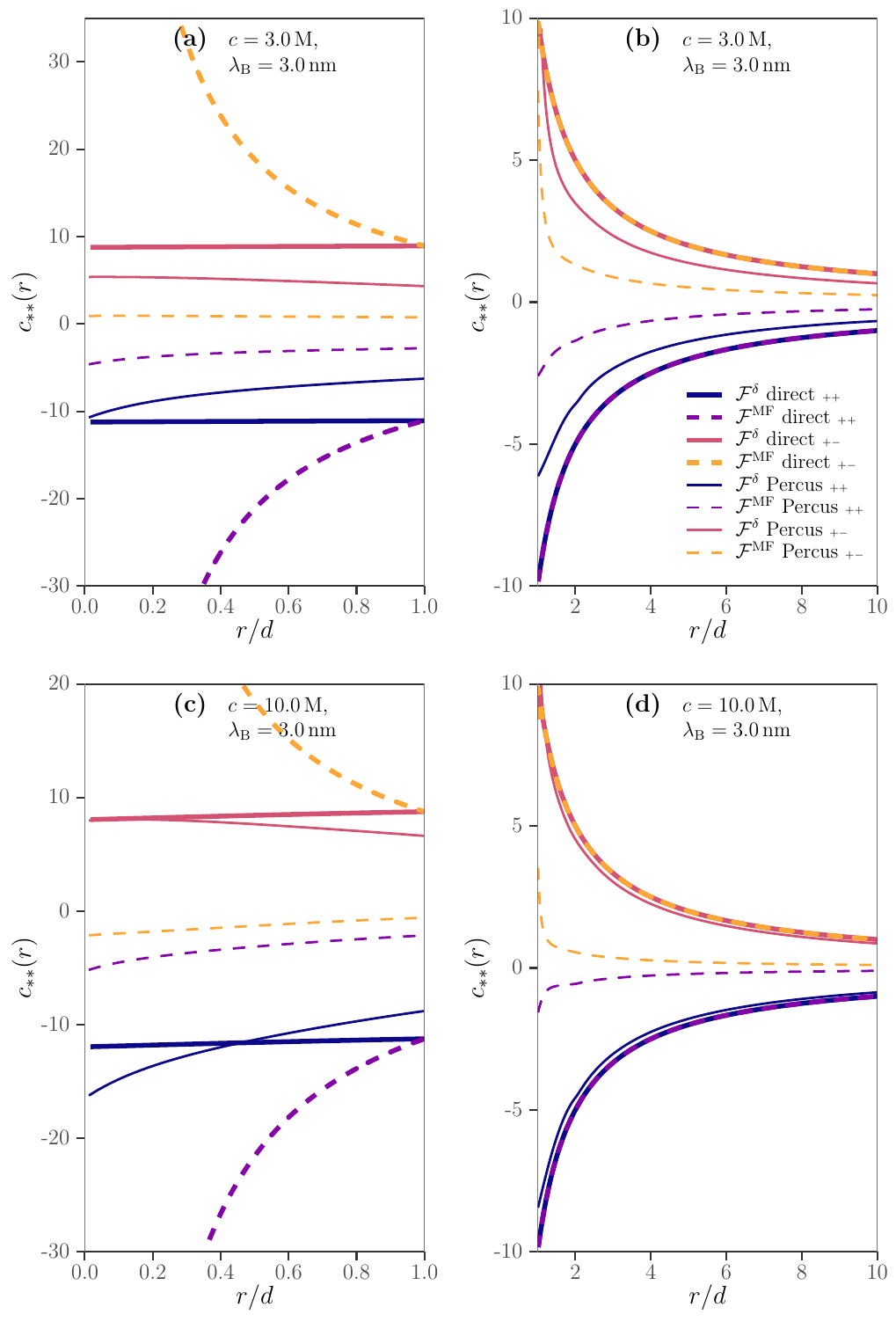}
\caption{
Two-body direct correlation functions of the \PM{} with diameters \hbox{$d:=d_{\scriptscriptstyle+}=d_{\scriptscriptstyle-}$} from the functionals $\functionaldelta$ (solid line) and $\functionalmf$ (dashed line) obtained via the second functional derivative of the excess free energy (bold lines) and the test-particle route (thin lines).
The \hbox{\raisebox{.15em}{\tiny$++$} and \raisebox{.15em}{\tiny$--$} correlations} are identical as well as the \hbox{\raisebox{.15em}{\tiny$+-$} and \raisebox{.15em}{\tiny$-+$} correlations}. 
The $++$ correlations (blue and purple) have negative values in all panels, while the $+-$ correlations (red and yellow) have positive values (except from the thin, dashed, yellow line in panel (c)).
To check the thermodynamic consistency, one can compare the curves of the same color.
The direct correlation function is split into the regions $r<d$ (panel (a) and (c)) and $r\geq d$ (panel (b) and (d)). Note the different range of the respective $y$-axis.}
\label{fig:directcorrelations}
\end{figure}

Accordingly, we now compare the direct correlation functions obtained via both routes for the functionals $\functionalmf$ and $\functionaldelta$, respectively. 
Our results are shown in \cref{fig:directcorrelations}. 
By using the difference between the direct correlation functions obtained via the two routes as a consistency measure \footnote{We ``measured'' consistency by a comparison of the shown curves, performed by the naked eye. For a rigorous definition of a measure, one can utilize a norm in function space.}, we find that the $\functionaldelta$ direct correlation functions are overall more consistent than the ones of $\functionalmf$.
For \hbox{$r<d$}, shown in panels (a) and (c), this is due to the diverging Coulomb potential which is contained in the second functional derivative of $\functionalmf$, while the second functional derivative of $\functionaldelta$ remains constant by construction (compare \cref{eqn:f-mf} and \cref{eqn:delta}).
For \hbox{$r\geq d$}, as shown in panels (b) and (d) of \cref{fig:directcorrelations}, the direct correlation functions from the second functional derivative are identical due to the definition of the pair potentials.
At both concentrations, the $\functionaldelta$ direct correlations are again much more consistent.
This is especially true for higher densities, where the $\functionaldelta$ functional also produces more accurate density profiles, as we discussed in~\cref{sec:results:profiles}.
%
\subsection{Sum rules}\label{sec:results:consistency}
%
Instead of comparing correlation functions as discussed in the previous subsection, we can also test certain sum rules to test thermodynamic consistency of our functional.
One of these sum rules is derived by a force balance argument for the hard planar wall and the particles pushing against it.
This so-called contact density theorem (\CDT{})~\cite{henderson1979exact} reads
\begin{equation}\label{eqn:cdt}
\beta P
=\sum_{\specA}\rho_{\specA}(z=R_{\specA})-\frac{\beta\sigma^2}{2\epsilon\epsilon_0}
\end{equation}
and relates the bulk pressure $P$ of the system to the contact densities at particle-wall contact and to the surface charge density $\sigma$ of the planar wall.
Whether it holds for a specific functional depends heavily on the functional's formulation and, thus, on its purpose.
Accordingly, some electrostatic functionals show thermodynamic inconsistencies by not satisfying the \CDT{}~\cite{kierlik1991density} and some functionals have been developed that satisfy it~\cite{roth2016shells}.

We test the \CDT{} by computing the three terms in \cref{eqn:cdt}.
Since the grand potential $\Omega = -PV$ is readily available in DFT, we can use it to calculate the bulk pressure. The surface charge $\sigma$ is given by the slope of the electrostatic potential $\Phi$ of the system at the walls, for instance for the system wall at $z=0$ via
\begin{equation}\label{eqn:surfacecharge}
\lim_{z\searrow0}\pdv{\Phi(z)}{z} = -4\pi\bjerrum\sigma \,.
\end{equation}
Both pressure and surface charge density are listed for selected system parameters in \cref{tab:parameterswalls} for the mean-field functional $\functionalmf$ and the restricted phase-space functional $\functionaldelta$.
The relative deviation between the left-hand side (\lhs{}) and the right-hand side (\rhs{}) of \cref{eqn:cdt} is calculated via \hbox{$|((\textrm{\lhs})-(\textrm{\rhs}))/(\textrm{\lhs})|$}.

The \CDT{} holds true for both the regular mean-field functional $\functionalmf$ and the core-corrected functional $\functionaldelta$.
There the given deviations are due to the finite numerical resolution of the grid the density profile is sampled on, which is indicated with a '$<$' sign in the table.
Thus the deviations might be even lower.
However, if the hard-sphere diameters differ between species, the deviations in the \CDT{} do not become smaller with increasing resolution as we found in additional calculations (results are not shown here).
Consequently, there is a real deviation of about $1\%$ in these cases.
We expect that this thermodynamic inconsistency will increase further when the inequality of hard-sphere diameters is increased.

Another approach to test thermodynamic consistency is via the virial pressure formula which for binary mixtures in bulk has the form
\begin{equation}\label{eqn:virialgeneral}
\beta P
=\sum_{\specA}\rho_{\specA} -\frac{2\pi\beta}{3}\sum_{\specA\specB}
\rho_{\specA}\rho_{\specB}\int_{0}^{\infty}\pdv{v_{\specA\specB}(r)}{r}
g_{\specA\specB}(r)r^3\inftes{r}\quad.
\end{equation}
If we supply this equation with the pair potential of the \PM{} from \cref{eqn:combinedpotential}, we can solve the integral of the hard-sphere potential and obtain
\begin{equation}\label{eqn:virialions}
\begin{split}
\beta P
=&\sum_{\specA}\rho_{\specA} +\frac{\pi}{12}\sum_{\specA\specB}
(\DADB)^3\rho_{\specA}\rho_{\specB}\,g_{\specA\specB}\kern-3pt\left(\DADBTwo\right)\\
&+\frac{2\pi\bjerrum}{3}\sum_{\specA\specB}
Z_{\specA}\rho_{\specA}Z_{\specB}\rho_{\specB}
\int_{0}^{\infty}rg_{\specA\specB}(r)\inftes{r} \,.
\end{split}
\end{equation}
Note, that the remaining integral does not diverge, because the constant part of the pair correlation functions cancel each other out due to the alternating sign in the sum over the valencies.

To evaluate the \lhs{} and the \rhs{} of \cref{eqn:virialions} we need to determine the pair-distribution function.
For this purpose, we use the density profiles from the test-particle route, because \citeauthor{archer2017standard}~\cite{archer2017standard} provide strong arguments for an increased accuracy of pair-correlation functions when using this route for mean-field functionals instead of the \OZ{} route.
In \cref{tab:parametersspherical} we summarize our results for systems of different concentrations and Bjerrum lengths.
The columns labelled ``Virial'' show the Virial pressure (\rhs{} of \cref{eqn:virialions}).
Clearly, the virial equation is not satisfied by any of the functionals, even though the \CDT{} showed thermodynamic consistency to a satisfying degree.
The overall agreement between the Virial pressure from simulation and from the Percus trick shows that the Percus trick produces more sensible results than the direct calculation via the grand potential.
For functionals with mean-field-like structure, this result is not surprising, because their electrostatic contributions vanish in bulk such that the obtained bulk pressure solely stems from the hard-sphere contribution to the functional. 
While the virial equation relies on the entire course of the pair-correlation functions, the \CDT{} only depends on contact quantities, which explains the difference in performance for both virial equation and \CDT{}.
In conclusion, testing the virial equation could be a good addition to a thorough analysis of functionals.

\section{Conclusion and Discussion}\label{sec:discussion}
In this work we derived a modified mean-field functional for the \PM{} of charged hard spheres.
For this purpose, we started from the Barker-Henderson perturbation theory, which, in first order, results in the standard mean-field electrostatic treatment.
For the higher order terms, we applied a low-density approximation, which resulted in a mean-field functional $\functionaltheta$ for a \emph{modified} electrostatic interaction potential where the center is cut out.
This electrostatic interaction potential is not continuous.
In a second modification we split the interaction potential into a repulsive hard-sphere contribution and an electrostatic, \emph{continuous} contribution, yielding a second functional $\functionaldelta$.
Both functionals contain correlation terms beyond the standard mean-field treatment which would be expected from the exact functional. 
In comparison to other sophisticated functionals for the \PM{}, for example the ones that are derived from the mean-spherical approximation (\MSA{}), our approach is built on the exact functional perturbation by approximating and neglecting certain terms.
Accordingly our approach is well related to the underlying Hamiltonian, transparent, and expandable.

To test our functionals we applied them to two systems of different geometry: a parallel plate capacitor and a test-particle setup to apply the Percus trick.
For comparison we also implemented the standard mean-field functional and performed \MD{} simulations.
Further, we consulted results from the literature for two \MSA{}-based functionals (MSAc and MSAu) that have been developed recently~\cite{roth2016shells,cats2021primitive}.
With these tools we studied density profiles and pair-correlation functions. 
The $\functionaltheta$ functional overestimates contact values greatly and overall performs weakly in predicting density profiles such that we did not investigated it any further.
Note that, nevertheless, it recently has been used to successfully explain a rapid switch in the experimentally measured oscillation length in the long-range decay of charge correlations~\cite{coupette2018screening}.
The $\functionaldelta$ functional, however, overall predicts density profiles better than the standard mean-field functional for all parameters we studied and, thus, is a true improvement beyond the standard mean-field treatment.
While both the mean-field and the modified functional agree with simulation results at low concentrations and weak electrostatic interactions, only the 
$\functionaldelta$ functional predicts the correct density profiles including layering effects at high concentrations. 
The dominating long-range correlations at low concentrations and high electrostatic interactions are accounted for by neither the $\functionalmf$ nor the $\functionaldelta$.
In particular due to \cref{eqn:lowdensity} and the approximation that the functional derivative of the involved total correlation function vanishes, long-range correlations were neglected in $\functionaldelta$.
However, the $\functionaldelta$ functional again shows good predictive capabilities if the ion concentrations are increased (\hbox{$>2\,\textrm{M}$}).

At uncharged walls one expects depletion effects due to the incomplete screening cloud of counterions around an ion that is close to the wall.
The only functional of those we tested that predicts this depletion is one of the \MSA{}-based functionals (MSAu in \cite{roth2016shells,cats2021primitive}) to which we compared our results.
This missing depletion is a drawback of mean-field like functionals (e.g.~$\functionalmf$,~$\functionaldelta$,~MSAc), because their electrostatic contributions vanish for equally sized ions as soon as the external electrostatic potential is turned off.
As a result, only the hard-sphere contribution to the functional remains, which leads to density profiles that do not show the expected depletion at the walls.
Similarly, bulk pressures are incorrect as a consequence to this missing contribution.
The situation would change if the interactions between like-charged and oppositely-charged ions were treated differently, for instance, in order to reduce mean-field repulsion between like-charged ions \cite{forsman2004cpb}.
Such changes can be implemented by correlation holes where interactions are cut below a certain threshold \cite{nordholm1984simple}, a concept also known from electronic structure calculations as exchange-correlation holes \cite{parr1994dft}. Different interaction ranges further can be respected in the concept of shells of charge \cite{jiang2021revisitShellsOfCharge} where, in contrast to our approach where only one of two charges is distributed on a spherical shell, the charge of each particle is smeared on some spherical shell.
Apart from this difference, our concept of modifying the electrostatic interaction potential inside the core can in general also be applied to implement different correlation holes.
Nevertheless, when we compared the $\functionaldelta$ functional to the \MSA{}-based functionals, the former predicts oscillations in the number and charge densities more accurate, while both show the correct decay behavior.
The mean-field functional, however, failed to capture the correct decay behaviour completely.
For the total correlation functions in bulk that we obtained via the Percus trick, we found similar results as those reported above for the density profiles.

In addition, we applied the Ornstein-Zernike equation to transform the total correlation functions into direct correlation functions such that we could compare them to the direct correlation functions obtained from a second functional derivative of the respective functionals.
Thus, we performed a direct check of the structural consistency of the functionals, where the $\functionaldelta$ functional performed much better.
Moreover, we tested the contact density theorem which we found to be valid numerically for both the modified functional $\functionaldelta$ and the mean-field functional $\functionalmf$ as long as the hard-sphere diameters are identical.
The virial equation for mixtures, that we tested as a second sum rule, however, is not satisfied.
To our knowledge the virial pressure comparison has not been done in other works presenting electrostatic functionals, although it seems to be a stricter measure in determining the quality of a functional.

Because of the advantages that the $\functionaldelta$ functional has over the regular mean-field functional, we think that it is the better choice in general. 
Further, the $\functionaldelta$ functional performs well in comparison to the \MSA{}-based functionals we consulted.
We suppose, that $\functionaldelta$ and the MSA functionals are comparable in terms of numerical speed, because the required operations (e.g. calculation of weighted densities and functional derivatives in FMT using Fourier transforms) are very similar. 
Of course, the computational speed is also affected by the numerical iteration scheme used for solving \cref{eqn:update}, for which, apart from the simple Picard iterations, sophisticated schemes like \cite{edelmann2016numerical} have been developed.

We discussed how the aforementioned drawbacks of functionals can be magnified by thermodynamic consistency checks which will help to tackle them systematically in future developments of functionals.
In this context we found that for increasing electrostatic interaction inconsistencies become more pronounced, because the bulk pressure (\lhs{} of \cref{eqn:virialions}) does not contain an electrostatic contribution due to the mean-field like structure of the functional.
Finally, improvements in terms of additional approximations from the Barker-Henderson theory might yield improved bulk values and better density profiles at uncharged hard walls, while keeping the underlying Hamiltonian \emph{within reach}.

\begin{acknowledgments}
We acknowledge support by the state of Baden-W\"{u}rttemberg through bwHPC
and the German Research Foundation (DFG) through Grant No. INST 39/963-1 FUGG
(bwForCluster NEMO) and through Project No. 406121234.
\end{acknowledgments}

\section*{Data Availability}
The data that support the findings of this study are available from the
first author upon reasonable request.

\appendix\label{sec:app}
\section{Bulk Derivatives}\label{app:thetaderiv}
%
As stated in \cref{sec:theory:funcderivs} the functional derivatives of $\functionaldelta$ from \cref{eqn:deltaderiv} are obtained by leaving out the last term of the functional derivatives of $\functionaltheta$ from \cref{eqn:thetaderiv} that read
\begin{equation}\label{eqn:app:thetaderiv}
\begin{split}
\fdv{\functionaltheta\funcdepend}{\rho_{\specA}(\vec{r})}&=
\frac{Z_{\specA}}{2\beta}\sum_{\specB}
\left(\phi_{\specB}\ast\tilde{\delta}_{\specA\specB}\right)(\vec{r})\\
&+\frac{Z_{\specA}}{2\beta}\sum_{\specB}\bjerrum\int\inftes{\vec{r}^{\,\prime}}
\frac{n_{\specA\specB}^{\delta}(\vec{r}^{\,\prime})}{|\vec{r}-\vec{r}^{\,\prime}|}\\
&-\frac{\bjerrum Z_{\specA}}{\beta}\sum_{\specB}n_{\specA\specB}^{\theta}(\vec{r})
\textrm{\,.}
\end{split}
\end{equation}
Thus, we show calculations just for the latter.
We start by setting the density profiles to a constant bulk value $\rho_{\specA}(\vec{r})=\rho_{\specA}$.
The first term of \cref{eqn:app:thetaderiv} yields
\begin{equation}
\begin{split}
&\frac{\bjerrum}{2\beta}Z_{\specA}
\sum_{\specB}Z_{\specB}\rho_{\specB}
\int\!\inftes{\vec{r}_2}\int\!\inftes{\vec{r}^{\,\prime}}
\frac{\delta\left(|\vec{r}^{\,\prime}-\vec{r}|-(\DADBTwo)\right)}
{\pi(\DADB)^2|\vec{r}_2-\vec{r}^{\,\prime}|}\\
&=
\frac{\bjerrum}{2\beta}Z_{\specA}
\sum_{\specB}Z_{\specB}\rho_{\specB}
\left(
\int_{\mathbb{R}^3\setminus\mathcal{V}^{\prime}}
\!\!\!\!\!\inftes{\vec{r}^{\,\prime}}\frac{2}{\DADB}
+\int_{\mathcal{V}^{\prime}}
\!\inftes{\vec{r}^{\,\prime}}\frac{1}{|\vec{r}^{\,\prime}-\vec{r}|}
\right)\\
&=
\frac{\bjerrum}{2\beta}Z_{\specA}
\sum_{\specB}Z_{\specB}\rho_{\specB}
\left(
\frac{\pi}{3}(\DADB)^2
+4\pi\!\!\!\int_{\DADBTwo}^{\infty}\!\!\!\!\inftes{r}r
\right)\\
&=
\frac{\bjerrum}{2\beta}Z_{\specA}
\sum_{\specB}Z_{\specB}\rho_{\specB}
\bigg(
\frac{\pi}{3}(\DADB)^2\\
&\hspace{9em}+
2\pi\lim_{r\rightarrow\infty}r^2-\frac{\pi}{2}(\DADB)^2
\bigg)\\
&=
-\frac{\pi}{12}\frac{\bjerrum}{\beta}Z_{\specA}
\sum_{\specB}Z_{\specB}\rho_{\specB}(\DADB)^2
\textrm{\,.}\\
\end{split}
\end{equation}
First, the shell theorem was applied.
Then the second integral can be solved trivially and gives a term depending on the hard sphere diameters.
Transitioning to spherical coordinates the first integral yields another term depending on the hard sphere diameters and a diverging term.
However, due to the imposed charge neutrality $\sum_{\specB}Z_{\specB}\rho_{\specB}=0$ this diverging term vanishes.

The second term of \cref{eqn:app:thetaderiv} is equivalent to the first one.
We see that the \rhs{} of the shell theorem in \cref{eqn:shelltheorem} only depends on $|\vec{r}_1-\vec{r}_2|$, which is symmetric in $\vec{r}_1$ and $\vec{r}_2$.
As a consequence, the equivalence follows via
\begin{equation}
\begin{split}
&\sum_{\specB}\left(\phi_{\specB}\ast\tilde{\delta}_{\specA\specB}\right)(\vec{r})\\
&=
\sum_{\specB}\bjerrum Z_{\specB}
\int\!\inftes{\vec{r}_2}\int\!\inftes{\vec{r}^{\,\prime}}
\frac{\rho_{\specB}(\vec{r}_2)\delta\left(|\vec{r}^{\,\prime}-\vec{r}|-(\DADBTwo)\right)}
{\pi(\DADB)^2|\vec{r}_2-\vec{r}^{\,\prime}|}\\
&=
\sum_{\specB}\bjerrum Z_{\specB}
\left(
\int_{\mathcal{V}^{\prime}}\!\inftes{\vec{r}_2}
\frac{\rho_{\specB}(\vec{r}_2)}{|\vec{r}_2-\vec{r}|}
+\int_{\mathbb{R}^3\setminus\mathcal{V}^{\prime}}\!\inftes{\vec{r}_2}
\frac{2\rho_{\specB}(\vec{r}_2)}{\DADB}
\right)\\
&=
\sum_{\specB}\bjerrum Z_{\specB}
\int\!\inftes{\vec{r}_2}\int\!\inftes{\vec{r}^{\,\prime}}
\frac{\rho_{\specB}(\vec{r}_2)\delta\left(|\vec{r}^{\,\prime}-\vec{r}_2|-(\DADBTwo)\right)}
{\pi(\DADB)^2|\vec{r}-\vec{r}^{\,\prime}|}\\
&=
\sum_{\specB}\bjerrum\int\inftes{\vec{r}^{\,\prime}}
\frac{n_{\specA\specB}^{\delta}(\vec{r}^{\,\prime})}{|\vec{r}-\vec{r}^{\,\prime}|}
\textrm{\,,}
\end{split}
\end{equation}
where the mentioned symmetry is utilized in the third and fourth line.
Thus, it remains to calculate the bulk case of the third term of \cref{eqn:app:thetaderiv},
\begin{equation}
\begin{split}
&-\frac{\bjerrum}{\beta}Z_{\specA}
\sum_{\specB}Z_{\specB}\rho_{\specB}
\int\!\inftes{\vec{r}^{\,\prime}}
\frac{2\,\theta(\DADBTwo-|\vec{r}^{\prime}-\vec{r}|)}{\DADB}\\
&=-2\frac{\bjerrum}{\beta}Z_{\specA}
\sum_{\specB}\frac{Z_{\specB}\rho_{\specB}}{\DADB}
\int_{\mathbb{R}^3\setminus\mathcal{V}^{\prime}}\!\inftes{\vec{r}^{\,\prime}}
\\
&=-\frac{\pi}{3}\frac{\bjerrum}{\beta}Z_{\specA}
\sum_{\specB}Z_{\specB}\rho_{\specB}(\DADB)^2
\textrm{\,.}\\
\end{split}
\end{equation}
Here, the integral yields the volume of a sphere.
Now, the modified chemical potential contributions, which are also found in \cref{eqn:mu-theta} and \cref{eqn:mu-delta}, follow as
\begin{equation}
\begin{split}
\mu_{\nu}^{\ast\theta}
&=\frac{\pi}{6}\frac{Z_{\specA}\bjerrum}{\beta}
\sum_{\specB}Z_{\specB}\rho_{\specB}(\DADB)^2\\
\mu_{\nu}^{\ast\delta}
&=-\frac{\pi}{6}\frac{Z_{\specA}\bjerrum}{\beta}
\sum_{\specB}Z_{\specB}\rho_{\specB}(\DADB)^2
\textrm{\,.}
\end{split}
\end{equation}
Note, that if the hard sphere diameters of all species are equal, the bulk derivatives vanish due to the charge neutrality $\sum_{\specB}Z_{\specB}\rho_{\specB}=0$ of bulk.
%
\section{Numerical Implementation}\label{app:numerical}
%
The numerical solution of Poisson equations is usually obtained by discretizing the difference quotient. Suppose that the potential $\Phi(r)$ of the test particle system is discretized into $n$ equidistant points in space with $\Phi(r_i = i\Delta{r}) = \Phi_i$ for $i = 1,\dots,n$.
The radial Poisson equation
\begin{equation}\label{eqn:radpoisson}
\frac{1}{r^2}\pdv{r}\left(r^2\pdv{\Phi(r)}{r}\right) = -4\pi\bjerrum q(r)
\end{equation}
of a charge distribution $q(r)$ can be discretized
\begin{equation}\label{eqn:radpoissonnum}
\frac{\Phi_{i-1} - 2\Phi_i + \Phi_{i+1}}{\Delta{r}^2}
+\frac{\Phi_{i-1}+\Phi_{i+1}}{r_i\Delta{r}}
=-4\pi\bjerrum q(r_i)
\end{equation}
into a system of linear equations.
$\Phi_{-1}$ and $\Phi_{n+1}$ are the two occurring boundary conditions to be specified.
They are subtracted from the equation to make them appear on the right-hand side.
Now, the system of linear equations can be represented by a matrix equation with a tridiagonal matrix, for which there are solving algorithms~\cite{press2007numerical} of linear numerical complexity $\mathcal{O}(n)$.
As boundary conditions we chose a combination of Dirichlet and Neumann conditions.
The Neumann condition is imposed at the center
\begin{equation}\label{eqn:radneumann}
\lim_{r\rightarrow0}\pdv{\Phi(r)}{r} = 0
\textrm{\,,}
\end{equation}
because we know that the derivative of the potential by the radial coordinate vanishes due to the spherical symmetry.
The outer boundary condition at $R_{\textrm{max}}$ is of Dirichlet type
\begin{equation}\label{eqn:raddirichlet}
\Phi(R_{\textrm{max}}) = \bjerrum\frac{Q_{\textrm{tot}}}{R_{\textrm{max}}}
\textrm{\,,}
\end{equation}
and is hence set according to Gauss' law.
Since all charges are contained within the sphere of radius $R_{\textrm{max}}$, the total charge $Q_{\textrm{tot}}$ is given by the integral of the total charge density over the volume of that sphere.
Note, that this quantity is usually non-zero, because the charge of the test particle is not included and its potential is incorporated into the external potential
\begin{equation}\label{eqn:radextpot}
V_{\textrm{ext},\specA}(r) = \bjerrum Z_{\specA}\frac{Q_{\textrm{test}}}{r}
\textrm{\,.}
\end{equation}
The Poisson equation for the hard wall geometry can be treated similarly as discussed above for the spherical geometry.
However, now we supply Dirichlet boundary conditions at both walls and set the potentials to $\pm\Delta\Phi/2$.
Hence, the external potential is already incorporated into the boundary conditions.

Now we compute the potential of the weighted charge densities $n_{\specA\specB}^{\delta}$
that occur in the functional derivatives of $\mathcal{F}_{\textrm{ex}}^{\theta}$ and $\functionaldelta$ in \cref{eqn:thetaderiv,eqn:deltaderiv}.
As we have seen previously the first two terms in both of these equations are identical and, for this reason, we only compute the second one.
Here, the potentials of the weighted charge densities are required.
To compute them, the system has to be extended such that the weighted charge densities are contained completely within the boundaries.
The new boundary conditions of the spherical system follow analogously to \cref{eqn:radpoisson,eqn:radpoissonnum,eqn:radneumann,eqn:raddirichlet,eqn:radextpot}.
For the planar system the electrostatic potential $\Phi(z)$ has to be extrapolated linearly (into the physical walls) to cover the whole extended system.
Hence, after computing the potential $\Phi$ of the physical system, the weighted charge densities are calculated to compute the corresponding species-specific potentials.

Fourier transforms are required for the convolutions in the \FMT{} functional and the restricted phase-space functionals $\functionaltheta$ and $\functionaldelta$.
In the spherical geometry the 3D Fourier transform ($\textrm{FT}$) boils down to 1D sine- and cosine-transforms ($\textrm{ST}$ and $\textrm{CT}$), for example
\begin{subequations}\label{eqn:sphericalft}
\begin{equation}
\begin{split}
\textrm{FT}\{\rho_{\specA}(r)\} &= \hat{\rho}_{\specA}(k) = \frac{4\pi}{k}\int_{0}^{\infty}\inftes{r}
r\rho_{\specA}(r)\sin(kr)\\
&= \frac{4\pi}{k}\textrm{ST}\{r\rho_{\specA}(r)\}
\textrm{\,,}
\end{split}
\end{equation}
\begin{equation}
\begin{split}
\textrm{FT}\{\hat{\rho}_{\specA}(k)\} &= \rho_{\specA}(r)
=\frac{1}{8\pi^3}\frac{4\pi}{r}\textrm{ST}\{k\rho_{\specA}(k)\}\\
\end{split}
\end{equation}
\end{subequations}
for the quantity $r\rho_{\specA}(r)$.
Numerical packages like \texttt{FFTW3}\cite{frigo2005fftw} also offer sine-~and cosine-transforms in addition to the Euclidean 1D transform.
Another way to compute STs and CTs is via Hankel transforms of order $\pm1/2$.
%
\section{Simulation Details}\label{app:simulation}
%
\begin{table}
\centering
\begin{tabularx}{\linewidth}{XXXXX}
\toprule
$c_{\textrm{bulk}}$ &
$c$ &
$\bjerrum$ &
$\Delta\Phi$ &
$L_x\!\times\!L_y\!\times\!L_z$\\
$[\textrm{M}]$ &
$[\textrm{M}]$ &
$[\textrm{nm}]$ &
$[\textrm{V}]$ &
$[\textrm{nm}^3]$\\
\midrule
0.1 & 0.108 & 0.726 & 0.1 & $20\times 20\times 15$\\
0.1 & 0.316 & 0.726 & 0.5 & $15\times 15\times 20$\\
0.1 & 0.109 & 4.25 & 0.1 & $50\times 50\times 12$\\
0.1 & 0.242 & 4.25 & 0.5 & $50\times 50\times 12$\\
2.0 & 1.930 & 4.25 & 0.1 & $15\times 15\times 12$\\
2.0 & 2.330 & 4.25 & 0.5 & $15\times 15\times 12$\\
5.0 & 4.870 & 4.25 & 0.1 & $10\times 10 \times 12$\\
5.0 & 5.145 & 4.25 & 0.5 & $10\times 10\times 12$\\
5.0 & 4.930 & 0.726 & 0.1 & $10\times 10\times 12$\\
5.0 & 5.368 & 0.726 & 0.5 & $10\times 10\times 12$\\
\bottomrule
\end{tabularx}
\caption{Parameters used for MD simulations of the \PM{} confined between two parallel charged plates.
The corresponding density profiles decay to a bulk concentration $c_{\textrm{bulk}}$, when an overall concentration $c=1.661\rho_{\scriptscriptstyle+}$ with $\rho_{\scriptscriptstyle+}=N_{\scriptscriptstyle+}/V$ is used.
The simulation box volume is given by $V=L_x\times L_y\times L_z$.}
\label{tab:simulation}
\end{table}
The molecular dynamics simulations for the density profiles for the planar geometry discussed in~\cref{sec:results:profiles} and for the radial distribution functions used in~\cref{sec:results:correlation} were carried out in the simulation package \texttt{ESPResSo~4.1.4}~\cite{weik2019espresso}.
As simulation units we use $\boltzmann T$ for energy, $3\cdot10^{-23}\,\textrm{g}$ for mass, $1\,\textrm{nm}$ for length, and $[\textrm{length}\sqrt{\textrm{mass}/\textrm{energy}}]\approx 2.7\,\textrm{ps}$ for time.
Instead of simulating the system at fixed chemical potentials $\mu_{\specA}$ in the grand ensemble ($\mu VT$) as used in \DFT{}, we prescribe fixed particle numbers $N_{\specA}$ in the canonical ensemble ($NVT$).
To ensure comparability between density profiles obtained from simulations and \DFT{}, we set the particle numbers such that the densities in bulk agree with each other.
The hard-sphere interactions are modelled by a steep, purely repulsive, truncated and shifted Lennard-Jones potential
\begin{equation}\label{eqn:lennardjones}
v_{\textrm{LJ}}(r) =
4\alpha\left[\left(\frac{r_{0}}{r}\right)^{12}
- \left(\frac{r_{0}}{r}\right)^6+c_\textrm{shift}\right]
\end{equation}
with an offset \hbox{$c_{\textrm{shift}}=\tfrac{1}{4}$}, a truncation radius of \hbox{$r_{\textrm{cut}}=d_{\specA}$}, a potential depth of \hbox{$\alpha=5\cdot10^3\,\boltzmann T$}, and \hbox{$r_{0}=2^{-1/6}d_{\specA}$}.
The hard-core interactions between walls and particles were implemented analogously.
The long-range electrostatic interactions are implemented via the P3M electrostatics solver, which employs a certain kind of Ewald summation method~\cite{weik2019espresso}.
The external electric field due to the charged walls is added via $4\pi\boltzmann T\bjerrum Q/e$, where $Q$ is the total charge of the walls.
The latter is obtained from the wall charge density of the \DFT{} calculations via~\cref{eqn:surfacecharge}.
The restricted symmetry in the simulations due to the hard walls is taken care of by an electric layer correction implemented in \texttt{ESPResSo}.

A preliminary equilibration run is performed such that the system reaches the expected average kinetic energy within a $1\%$ accuracy.
Afterwards several particle trajectories are recorded by taking snapshots every $300$ time steps until in total $10^5$ snapshots are collected.
These snapshots are then processed into a histogram with $400$ bins.
The simulation box dimensions depended on the concentration of the simulated system and on how far the density profiles were expected to decay into the bulk.
The parameter combinations for all reported systems are presented in~\cref{tab:simulation}.

%
\end{document}